 \documentclass[twocolumn]{aastex62}
\usepackage{enumitem}
\shorttitle{Betelgeuse as a Merger}
\shortauthors{Chatzopoulos et al.}
\begin{document}

\title{IS BETELGEUSE THE OUTCOME OF A PAST MERGER?}

\correspondingauthor{Emmanouil Chatzopoulos}
\email{chatzopoulos@phys.lsu.edu}

\author[0000-0002-0786-7307]{E. Chatzopoulos}
\affiliation{Department of Physics \& Astronomy, Louisiana State University, Baton Rouge, LA, 70803, USA}
\affiliation{Hearne Institute of Theoretical Physics, Louisiana State University, Baton Rouge, LA, 70803, USA}

\author{Juhan Frank}
\affiliation{Department of Physics \& Astronomy, Louisiana State University, Baton Rouge, LA, 70803, USA}

\author{Dominic C. Marcello}
\affiliation{Department of Physics \& Astronomy, Louisiana State University, Baton Rouge, LA, 70803, USA}
\affiliation{Center for Computation and Technology, Louisiana State University, Baton Rouge, LA, 70803, USA}

\author{Geoffrey C. Clayton}
\affiliation{Department of Physics \& Astronomy, Louisiana State University, Baton Rouge, LA, 70803, USA}



\begin{abstract}

We explore the possibility that the star $\alpha$~Orionis (Betelgeuse) is the outcome of a merger that occurred in
a low mass ratio ($q = \mathcal{M}_{\rm 2}/\mathcal{M}_{\rm 1} =$~0.07--0.25) binary system 
some time in the past hundreds of thousands of years.
% several thousand years ago. 
To that goal, we present a simple analytical model to approximate the perturbed internal structure of
a post--merger object following the coalescence of a secondary in the mass range 1--4~$M_{\odot}$ into the envelope of a 15--17~$M_{\odot}$
primary. We then compute the long--term evolution of post--merger objects for a grid of initial conditions and make predictions
about their surface properties for evolutionary stages that are consistent with
the observed location of Betelgeuse in the Hertzsprung-–Russell diagram. 
We find that if a merger occurred after the end of the primary's main--sequence phase,
while it was expanding toward becoming a red supergiant star and typically with radius $\sim$~200--300~$R_{\odot}$, then it's envelope
is spun--up to values which remain in a range consistent
with the Betelgeuse observations for thousands of years of evolution. We argue that the best
scenario that can explain both the fast rotation of Betelgeuse and its observed large space velocity is one where a binary
was dynamically ejected by its parent cluster a few million years ago
and then subsequently merged. An alternative scenario in which the
progenitor of Betelgeuse was spun up by accretion in a binary and
released by the supernova explosion of the companion requires a finely tuned set of conditions but cannot be ruled out.

\end{abstract}



\section{Introduction} \label{intro}

Recent observational surveys of massive main--sequence stars  have shown that a large number of them
rotate rapidly with some reaching rotational speeds as high as $\sim$~450~km~s$^{-1}$ 
and with a bimodal distribution of rotational velocities \citep{2011ApJ...743L..22D,2013A&A...550A.109D,2013A&A...560A..29R}.
In addition, the {\it Kepler} space mission reports observations of 17 rapidly--rotating $<$~$\sim$~1--3~$M_{\odot}$ 
stars in the giant phase of their evolution with rotational speeds up to $\sim$18 times that of the Sun (\citealt{2015ApJ...807L..21C}; see
also studies of the rotation rates of young spectroscopic binaries by \citealt{2015A&A...580A..92R}).
These findings, coupled with the obsevation that the majority ($\sim$~60\%) of massive stars are members of binary stellar systems
\citep{2009AJ....137.3358M,2015A&A...580A..93D} and that most of the massive stars in such 
systems will at sometime undergo binary interactions
with a third of them even experiencing a merger \citep{2012Sci...337..444S}, 
indicate the importance of studying the effects of stellar mergers on the long--term evolution
of the rotational properties of massive stars.

Binary population synthesis models presented by \citet{2013ApJ...764..166D,2014ApJ...782....7D} support the idea
that $\sim$~20\% of massive main--sequence stars are the products of binary interaction. Furthermore, the presence of a
low velocity component of the bimodal distribution of the rotational velocities of massive stars \citep{2013A&A...550A.109D}
together with evidence of strong magnetism in some massive stars \citep{2012MNRAS.426.2208G,2016MNRAS.457.2355S} 
motivated theoretical studies of the merger of a massive star with a lower--mass companion as an avenue to produce these properties
\citep{2009MNRAS.400L..71F,2012ARA&A..50..107L}. Despite all of the evidence pointing toward the prevalence
of stellar mergers and their subsequent effects on the evolution of massive stars, detailed long--term evolution
studies of such post--merger stars are limited \citep{1990A&A...227L...9P,1992ApJ...391..246P,2017MNRAS.469.4649M}.

The massive red supergiant star $\alpha$~Orionis, popularly known as Betelgeuse, is a potentially interesting
candidate for a past binary merger event, partly because of its apparent high--rotational velocity.
Using long--slit spectroscopy across the minimally resolved disk of Betelgeuse 
obtained by the {\it Hubble Space Telescope} the estimated surface rotational velocity of the supergiant star
($v_{\rm rot} \sin(i)$) is $\sim$~5~km~s$^{-1}$ \citep{1996ApJ...463L..29G,1998AJ....116.2501U}. Recent observations by the {\it Atacama
Large Millimeter Array} ({\it ALMA}) appear to further support this result even within the uncertainties imposed by large--scale convective
motions on the star's surface \citep{2017MNRAS.465.2654W,2018A&A...609A..67K}. Furthermore, the observation of enhanced
nitrogen on the surface of Betelgeuse is indicative of enhanced mixing, triggered by rotation \citep{2013EAS....60...17M}. In particular
the measured N/C (nitrogen to carbon) and N/O (nitrogen to oxygen) surface abundance ratios for Betelgeuse are 2.9 and 0.6 respectively
while standard solar values are N/C$=$0.3 and N/O$=$0.1 \citep{1984ApJ...284..223L}.
High rotation during the supergiant phase is not found in stellar evolution calculations of single massive stars -- including those that are rapid 
rotators at the Zero Age Main Sequence (ZAMS) -- nor is expected by
simple arguments of angular momentum conservation \citep{2017MNRAS.465.2654W}. Measurements of giant and supergiant star rotation
rates support this argument \citep{2017A&A...605A.111C}.

In addition to being a rapid rotator, Betelgeuse is also a known runaway star with a measured space velocity of $\sim$~30~km~s$^{-1}$
and a kinematic age of $\sim$~7--11~Myr \citep{2008AJ....135.1430H,2017AJ....154...11H}. Backwards extrapolation of Betelgeuse's trajectory
has led some to suggest that its possible birthplace is the Orion OB1a association \citep{2005AJ....129..907B} yet alternative scenarios involving
two dynamical kicks have been suggested \citep{2008hsf1.book..459B}. Betelgeuse's flight through the interstellar medium is also illustrated
by {\it HST} observations of a bow shock forming a  0.14~$M_{\odot}$ swept--up shell of material around the star at a radius of $\sim$~6--7~arcmin
corresponding to a physical distance of $\sim$~0.8~pc using the measured $\sim$~400 pc distance of Betelgeuse \citep{1997AJ....114..837N}.
The morphology of this structure is attributed to wind from the star sweeping up interstellar medium in the direction of motion 
\citep{2012A&A...541A...1M,2014arXiv1406.0878M}.

To investigate the long--term effects of a merger during the post--main sequence evolution of massive stars and to model the observed
properties of Betelgeuse in particular, simulations spanning a variety of time--scales that involve diverse numerical approaches
are necessary: the merger occurs in a dynamical time--scale while the post--merger thermal adjustment and susequent evolution
in thermal and nuclear time--scales. Furthermore, any successful model of Betelgeuse has to account for both its
rapid rotation rate and its runaway nature, so a sequence of events (dynamical ejection of a binary followed by a merger) have
to be invoked.

In the present work we compute an approximate, analytical model of the perturbed, post--merger structure of a massive
star assuming that it suffered the coalescence of a smaller, 1--4~$M_{\odot}$ secondary into its envelope, and compute the long--term
evolution of the post--merger object under the assumption of different initial conditions for the original binary system. 
We find that a merger of a binary with mass--ratio in the range $q =$~0.07--0.25 occurring during the ``Hertzsprung gap''--crossing phase, 
after the end of the main--sequence, can yield high surface equatorial rotational velocities in agreement with observations of
Betelgeuse. In addition, our models predict that high rotation is preserved for hundreds of thousands of years following the merger event.
We also show that such runaway low--mass ratio binary systems can be dynamically ejected from clusters without being disrupted, allowing
for the merger to occur at a later time.

Our paper is organized as follows: In Section~\ref{merger_necessary}
we present a qualitative analysis of the kinds of binaries likely to
reproduce the observed properties of Betelgeuse considering both the
merger and the accretion scenarios and find the merger path more
compelling. However, we are not able to rule out completely the accretion option given the current uncertainties in the evolution of massive binaries. In Section~\ref{merger_analytical} 
we present our initial setup and synchronization conditions for low mass--ratio binaries of interest and introduce a simple analytical model 
to approximate the post--merger structure of the primary. In addition, we present a 3D hydrodynamical simulation of a merger to be used for benchmarking against some of our
analytical assumptions. The {\it MESA} evolution calculations for the post--merger objects
are presented in Section~\ref{evol}. Section~\ref{results} presents the results of our calculations including
comparisons with the observed properties of Betelgeuse. Finally in Section~\ref{disc} we discuss the implications of our
results for Betelgeuse as well as other observations and describe our plans for future work.

\section{Spun--up by Accretion or Merger?}\label{merger_necessary}

Betelgeuse is not only a fast rotating supergiant but it is also a runaway star, probably born in one of the clusters or 
sub-associations of OB1 in Orion \citep{2008AJ....135.1430H,2017AJ....154...11H}. In this section we consider a variety of initial progenitor configurations and evolutionary histories
for Betelgeuse to show, using angular momentum (AM) conservation and simple dynamical arguments, that it
is very difficult to reconcile both the large space velocity and fast rotation without invoking a recent binary interaction 
yielding a fast rotating single progenitor that subsequently evolved within a million years or so to the red supergiant we see today.
We consider two possible scenarios: a binary merger, preferably in a post--MS stage of evolution, or a late case B accretion event followed by a supernova explosion of the donor.
The high space velocity suggests ejection from
a binary as a result of a supernova (SN) explosion of a much more massive companion that disrupted the binary \citep{1961BAN....15..265B}, or a dynamical ejection 
from a cluster or compact OB association \citep{1964Natur.202.1319P}. 

If Betelgeuse had evolved as a single star, given its present red supergiant (RSG) status, its age would be around 8-10 million years.
Early ejection as a single star either by the disruption of a cluster binary or dynamical escape from a cluster are unlikely
to yield a rapid rotator in the present supergiant stage. Single massive stars lose a fraction $f$ of their mass which carries away some AM through winds during the main sequence (MS) phase.
In particular, O stars evolve through rapid mass and AM losses to much slower rotating B stars with $v \sin i \leq$~50~km~s$^{-1}$ (\citealt{rotation,2019A&A...622A..50H} and references therein). 
During their MS evolution, it is reasonable to assume that stars remain close to solid body rotation. For example, for a
20~$M_{\odot}$ star on the ZAMS, with radius $R_{\rm i}$ and gyration radius $\beta_{\rm i}$,  its initial AM can be written as $J_{\rm i} = \beta_{\rm i}^2\mathcal{M}_{\rm i} R_{\rm i} v_{\rm eq,i} $.
Similarly, after losing a mass fraction $f$ of the initial mass, $\mathcal{M}_{\rm f} = (1-f) \mathcal{M}_{\rm i}$, its AM on the Terminal Age Main Sequence (TAMS) can be written as
$J_{\rm f} = \beta_{\rm f}^{2}(1-f)\mathcal{M}_{\rm i}R_{\rm f} v_{\rm eq,f}$. Without a detailed model for the mass and AM losses carried by the wind we do not know a priori the average specific
AM lost to the wind. We may expect it to be on the order of the specific angular momentum of the surface layers, and we can write in general that the total AM lost to the wind is:
\begin{equation}
J_{\rm f} - J_{\rm i} = - f  \mathcal{M}_{\rm i} \eta R_{\rm i} v_{\rm eq,i},
\end{equation}
where $f  \mathcal{M}_{\rm i}$ is the mass loss, and $\eta R_{\rm i} v_{\rm eq,i}$ is the average specific AM lost. Using AM conservation and solving for the equatorial velocity at the TAMS, we get
\begin{equation}
v_{\rm eq,f}  = v_{\rm eq,i} \frac{\beta_{\rm i}^2 (1- f\eta) R_{\rm i}}{\beta_{\rm f^2} (1-f) R_{\rm f}},
\end{equation}
where the initial equatorial velocity is unlikely to exceed $\sim$~200~km~s$^{-1}$ at the ZAMS. One can easily see that for the example considered, $R_{\rm i} \sim$ 6~$R_{\odot}$, $R_{\rm f} \sim$~18~$R_{\odot}$
and values for the gyration radii $\beta_{\rm i} \simeq$~0.29 and $\beta_{\rm f} \simeq$~0.16 \citep{1989A&AS...81...37C}, it is relatively easy to decrease $v_{\rm eq,f}$ to 50~km~s$^{-1}$ and below.
For a fixed value of $f$, $v_{\rm eq,f}$ decreases linearly with $\eta$. Despite its simplicity, this model suggests that you need at least a mass loss fraction of $f\geq$~0.07
to reduce the equatorial velocity at the TMS to $\sim 50$~km/s, unless one appeals to magnetic effects that could make $\eta>1$. During the rapid expansion while the post--main sequence (PMS) star
crosses the Hertzsprung Gap we may assume that redistribution of AM is inefficient and that a better approximation is to take ``local'' AM conservation and write $v_{\rm eq} = v_{\rm f}R_{\rm f}/R$, where the
subscript ``f'' stands for values at the TAMS, while $v_{\rm eq}$ and $R$  represent the values attained when the star has expanded to RSG size. With $v_{\rm f}\approx 50$ km/s,
$R_{\rm f} \approx 18$~$R_{\odot}$ and $R\sim 10^3$~$R_{\odot}$ this estimate yields $v_{\rm eq,f} \sim$~1~km~s$^{-1}$. This this is very much an upper limit. In fact, detailed simulations of the evolution of single massive stars,
including mass and AM losses, from ZAMS to the supergiant stage, typically yield $v_{\rm eq,f} <$~1~km~s$^{-1}$
\citep{2008A&A...489..685E,2012A&A...537A.146E,2011A&A...530A.115B,2011A&A...530A.116B,2017MNRAS.465.2654W}.
If Betelgeuse's progenitor was initially the less massive member of a massive binary, then we need to consider how the
spin--up by accretion would impact the above argument.  We discuss this possibility  later in the context of the current incomplete understanding
of the complex evolution of massive binaries of arbitrary initial mass ratios and separations.

A further apparent difficulty with Betelgeuse is that a simple backward extrapolation of its known space velocity 
does not appear to bring it close to any plausible sub--association of OB1 as its birth place \citep{2008hsf1.book..459B}. 
This suggests a two step process: 1. a dynamical ejection of a binary within the first few million years of Betelgeuse's 
birth cluster, and 2. a subsequent merger of the binary or a supernova explosion of the more massive component, 
releasing the surviving Betelgeuse at some post--MS stage of its evolution.  
Recent binary stellar evolution models suggest that the SN mechanism alone is not
as efficient in producing runaway stars as it was previously thought, with only a tiny fraction of stars ejected by
the SN of their companions reaching velocities $>$~30~km~s$^{-1}$ \citep{2019A&A...624A..66R}. A two--step
process gets around this difficulty. It is, however, noted that for the merger to change the flight direction of the post--merger
it would require significant asymmetric mass loss. This is possible, (e.g. mass loss via the $L_{\rm 2}$ point), but that would also
lead to the removal of some AM. A SN explosion in the binary instead would eject the companion star with a direction depending on the
orbital phase of the binary, which requires fine tuning to be aligned with the previous space velocity in the double ejection scenario.

For a binary to remain bound after dynamical ejection due to interactions with other binaries and members
of the early cluster, its binding energy must be greater than the kinetic energy of the escaping binary plus the
binding energy of the original binary to the cluster:
\begin{equation}
\frac{G \mathcal{M}_{\rm 1} \mathcal{M}_{\rm 2}}{2 a_{\rm i}} > \frac{1}{2} \mathcal{M} (v_{\rm ej}^{2}+\sigma_{\rm c}^{2}),
\end{equation}
where $\mathcal{M} = \mathcal{M}_{\rm 1} + \mathcal{M}_{\rm 2}$ is the initial total mass of the binary, $v_{\rm ej}$ is the
ejection velocity of the binary, and $\sigma_{\rm c}$ is the central velocity dispersion in the cluster.
We take $v_{\rm ej}$ to be on the order of the observed runaway velocity of Betelgeuse ($\sim$~30~km~s$^{-1}$),
and $\sigma_{\rm c} \simeq$~6.5~km~s$^{-1}$, typical for a cluster with mass $10^{4}$~$M_{\odot}$ (in agreement with the
measured mass of the Ori OB1 association, the proposed birthplace for Betelgeuse; \citealt{1964ARA&A...2..213B,1994A&A...289..101B}) 
and half--mass radius of $\sim$~1~pc. Thus the dominant term is the runaway kinetic energy and we neglect $\sigma_{\rm c}$ in the following.
From the above argument we conclude that the initial separation should not exceed:
\begin{equation}
a_{\rm i} < \frac{G q \mathcal{M}}{(1+q)^{2} v_{\rm ej}^{2}}, \label{Eq:amax}
\end{equation}
where $q = \mathcal{M}_{\rm 2}/\mathcal{M}_{\rm 1}$ is the mass ratio of the binary. 
This condition turns out not to be very restrictive
because the components of massive binaries with  $\mathcal{M}\sim 30 M_\odot$ and initial separations greater than 15-20 AU  would evolve as single stars without interacting.  Therefore, most of the binaries
considered in the analysis that follows would survive the ejection from their parent cluster. This is further
supported by results of $N$--body simulations of the dynamical cluster evolution focusing on binary--binary interactions
predicting a considerable fraction of binaries that escape the cluster without being disrupted 
\citep{1983MNRAS.203.1107M,1986ApJS...61..419G,2016A&A...590A.107O,2018A&A...612A..74K,2019MNRAS.484.1843W}. 
In particular, the simulations in \citet{2016A&A...590A.107O} (and private communication with Dr. Pavel Krupa) show that for a variety of initial assumptions, 
a handful of binaries with $q<0.3$ and orbital periods in the range $2<\log{P_{\rm days}}<3$ are ejected in the first few million years of cluster life. 
As we shall see, these are precisely the kinds of binaries whose merger has the potential to account for the properties of Betelgeuse.

Without loss of generality we may assume that $\mathcal{M}_{\rm 1} >\mathcal{M}_{\rm 2}$ or $q \leq$~1 for the ejected binary. 
We further assume that one or both components are moderately massive, such that $\mathcal{M} \simeq$~15--20~$M_{\odot}$. 
The evolution of massive binaries is subject to many uncertainties and there are significant differences of detail in the 
outcomes of such simulations by various groups. Mass transfer starts as the more massive 
component evolves first and overflows its Roche lobe. Apart from differences of detail and uncertainties in the exact critical
values of $q$, most investigators distinguish the following regimes: 
If the binary components are similar in mass, evolution to a contact configuration is avoided. If $q_{\rm cont} < q < 1$, where $q_{\rm cont}$, which depends on the evolutionary phase of the donor, is somewhere in the range (0.65, 0.8), $\mathcal{M}_{\rm 1}$ overflows its Roche lobe, mass
transfer begins either in Case A or early Case B \citep{1967ZA.....65..251K}, and is high but stable, leading to inversion of the mass ratio and an
``Algol" analogue, with separation and period increasing as the system evolves \citep{1994A&A...290..119P,2010NewAR..54...39P}. While the ultimate fate of such
binaries is of interest elsewhere, they do not produce Betelgeuse--like stars. As an example of such evolutions, 
a binary consisting of a 16 $M_\odot$ primary and a 14 $M_\odot$ secondary ($q=0.875$) evolves through Case A 
and Case AB mass transfer
dumping most of the H-rich primary envelope onto the secondary yielding a fast rotating blue supergiant (BSG), while
the primary becomes a He star. In many cases, especially if the mass transfer is conservative, the secondary explodes as a SN first \citep{1994A&A...290..119P,2001A&A...369..939W,2015ApJS..220...15P}.

For mass ratios close to $q_{\rm cont}$, the outcome is dependent on the initial orbital period $P_{\rm i}$. For example, \citet{2013ApJ...764..166D} investigated in detail the outcomes of the evolution of a
binary of $q=0.75$, with initial components 
$\mathcal{M}_{\rm 1}=20 M_\odot$ and $\mathcal{M}_{\rm 2}=15$~$M_{\odot}$. For $P_{\rm i}<2$ days, after Case A mass transfer, a rapidly rotating merger resulted which spun down quickly.
For $2<P_{\rm i}<5$ days, after Case AB or even Case ABB mass transfer they obtained a rapidly spinning secondary which then also spins down in a couple of million years. For even longer orbital periods with 
10 d $<P_{\rm i} < 10^3$ d, mass transfer sets in as the primary is evolving to become a red giant (Case B) and that also may lead to a rapidly rotating secondary, now more massive than the
original primary. However, the details of the evolution depend on metallicity, the efficiency of accretion (the faction of mass transfer actually captured by the secondary), mass loss via winds,
and the efficiency of semiconvection. Sometimes the H-rich accreted material is mixed down into the core and a BSG results, sometimes mixing is not so efficient and a red supergiant (RSG)
with a small core results. In addition, it turns out that which of the two stars goes SN first also depends on these details \citep{2001A&A...369..939W}.
Therefore, in order to get a single RSG whose companion explodes first as a SN releasing a spun-up star with properties similar to Betelgeuse may
require a narrow set of conditions, but to estimate the probability for this channel would require a series of binary stellar evolution calculations
specifically designed for this purpose, perhaps starting with an ad hoc set of binary population synthesis. This is beyond the scope of the present
work, in which we focus on the merger channel recognizing that there may well exist a set of binary parameters that would result in Betelgeuse-like
outcomes in which accretion yields the relatively high equatorial velocity observed. At this point we cannot rule that out completely.

For lower mass ratios, down to $q_{\rm blue}$, where $q_{\rm blue}$ 
is somewhere in the range (0.25, 0.33), $\mathcal{M}_{\rm 2}$ is unable to adjust to the mass transfer, expands and the system evolves into a massive
contact binary. As time proceeds, the contact binary shrinks converting orbital AM into circulation or rotation and ultimately 
merges producing a massive analog of a ``blue straggler". This occurs because during the merger hydrogen--rich material gets 
mixed into the core rejuvenating the merged object. While this may well produce rapidly rotating BSGs, like the presumed 
progenitor of SN 1987A \citep{1990A&A...227L...9P,2019MNRAS.482..438M}, they are not applicable to Betelgeuse. 

Finally, for even lower mass ratios, we have the cases that we consider in more detail in subsequent sections which in our view offer a more likely pathway to a star with Betelgeuse's properties.
For mass ratios $q < q_{\rm blue}$, and $P_{\rm i} >$ few tens of days, mass transfer starts as an early Case B, after H is exhausted in the primary,
is rapid and results in the primary envelope engulfing the much-lower mass secondary, which then spirals inward and producing a merger.
In this scenario, the primary's helium core is surrounded by a H--burning shell. When the secondary
reaches the critical tidal disruption distance from the core of the primary, a tidal stream will form transporting fresh H fuel toward the
core as described in \citet{2002PhDT........25I,2002MNRAS.334..819I,2003fthp.conf...19I,2016MNRAS.457.2355S}. The penetration depth of the stream into
the core of the primary determines the extent of its rejuvenation; if fresh fuel reaches the core then core H--burning will be
re--ignited and the star may evolve toward the BSG phase. If, on the contrary, the stream does not penetrate deep into the core but
rather mixes--in with the H--burning shell then the star will keep evolving toward the RSG stage. Using the same arguments as
the ones presented in \citet{2002MNRAS.334..819I} involving the Bernoulli integral to calculate the penetration depth, we have found
that in none of our models the stream is expected to penetrate the core of the primary thus suggesting a post--merger evolution
toward the RSG phase. This is in agreement with the ``quiet merger'' scenario discussed in \citet{2003fthp.conf...19I} that, in addition,
predicts the formation of rapidly rotating supergiants via this channel. We have also confirmed this via a 3D numerical merger simulation
for a system with parameters in our range of interest (a 16~$M_{\odot}$ and a 1~$M_{\odot}$ secondary; Section~\ref{OctoTiger3D}).
When $q < q_{\rm blue}$, the lower mass companion is engulfed in the atmosphere of $\mathcal{M}_{\rm 1}$ as it overflows its Roche lobe and $\mathcal{M}_{\rm 2}$ 
ends up spiralling inward and is tidally disrupted outside 
the core of the primary, so not much hydrogen--rich material is mixed into the core and the final result is a red supergiant that has a high
equatorial velocity as these layers absorbed the bulk of the orbital angular momentum of the secondary. The amount of orbital angular 
momentum depends mostly on the separation of the binary when the primary overflows its Roche lobe. These are the cases most likely applicable to Betelgeuse.

\begin{figure}
\begin{center}
\includegraphics[angle=0,width=9cm]{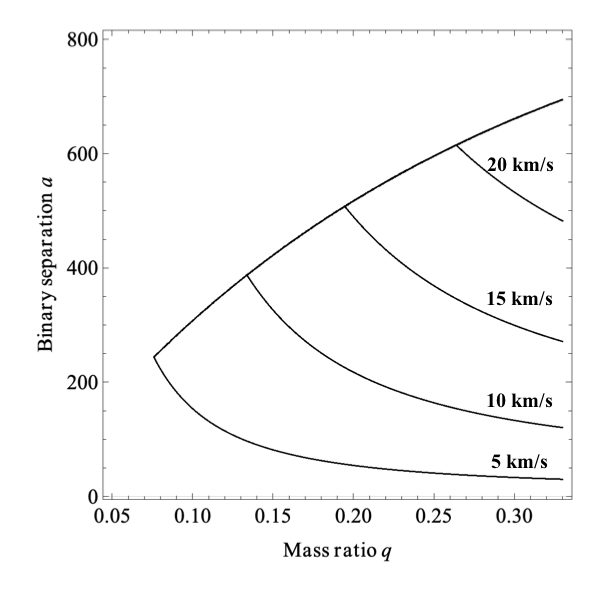}
\caption{Contours of final equatorial velocity of the merged star at the red supergiant stage.
For the purposes of the estimate here we assumed a total mass $M =$~20~$M_{\odot}$, and $R_{\rm f} =$~1000~$R_{\odot}$. The
binary separation is in units of solar radii and the contours, from bottom to top, correspond to 5, 10, 15 and 20~km~s$^{-1}$.
The top boundary is the maximum separation allowed by Equation~\ref{Eq:constrain}.}
\label{Fig:survivability}
\end{center}
\end{figure}

Which binary parameters subject to the restrictions described above are likely to yield an equatorial velocity exceeding $\sim$~5~km~s$^{-1}$ 
when the merged object evolves into a red supergiant? Before attempting to answer this question in the following sections, it is worth 
obtaining a simple estimate based on angular momentum conservation. Most of the angular momentum of the
pre--merger binary is in the orbital angular momentum $J_{\rm orb} = \mu a  v_{\rm orb}$, where $\mu$ is the reduced mass, $a$ is the
binary separation, and $v_{\rm orb} = (G \mathcal{M}/a)^{1/2}$ is the Keplerian velocity for a circular orbit. Numerical hydrodynamic simulations of 
binary mergers show that the immediate post--merger object is roughly of the same size as the original binary and is differentially rotating.
Nonetheless, for our present purposes, we estimate the equatorial velocity of the merged object by assuming a uniform density sphere 
rotating as a solid body, and we set $J_{\rm orb} = (2/5) \mathcal{M} a v_{\rm pm}$, where $v_{\rm pm}$ is the post--merger equatorial velocity.
Since in all the cases of interest, contact, mass transfer, and merger occur during the crossing of the Hertzsprung Gap, 
the subsequent evolution to red supergiant is very rapid, we may assume ``local" angular momentum conservation and set 
$v_{\rm eq} \simeq v_{\rm pm} a/R_{\rm f}$. Therefore we have as our order--of--magnitude estimate the following result:
\begin{equation}
v_{\rm eq} (q,a) = \frac{5 q}{2 (1+q)^{2}} v_{\rm orb} (\mathcal{M},\alpha) \frac{\alpha}{R_{\rm f}}.\label{Eq:constrain}
\end{equation}
Figure~\ref{Fig:survivability} shows the binary parameter region likely to yield equatorial velocities consistent with the measured values
of $v_{\rm eq} \sin i$ for Betelgeuse, for an assumed initial binary mass of 20~$M_{\odot}$. All contours are limited at the maximum sepration 
allowed for a binary to survive ejection as estimated by Equation~\ref{Eq:amax}. We will explore in detail
the outcomes of mergers with such initial parameters in Sections~\ref{evol} and ~\ref{results}.

\section{Merger Model}\label{merger_analytical}

\subsection{{\it Initial Conditions for the Binary System}}\label{assump}

We begin by considering a few possible initial configurations for our binary  differing only in the separation (and orbital period) at the point 
at which the primary fills its Roche lobe initiating mass transfer and a rapid evolution leading to the plunge of the 
secondary into the envelope of the primary. These distinct binary separations result in contact at different stages of evolution of the primary through the
Hertzsprung Gap. We will consider 4 situations corresponding to the primary radius $R_1$ filling its Roche lobe when $R_{\rm 1} =$~12~$R_\odot$,
100~$R_\odot$, 300~$R_\odot$, and 500~$R_\odot$. These situations correspond to contact soon after the end of the main sequence phase,
two intermediate stages in the middle of the crossing, and one near the red end of the Hertzsprung Gap as the ascent of the red giant branch
is to begin shortly. 
 
The total angular momentum of such a binary consists of the orbital and the spin angular momenta. As the primary evolves, initially tides will 
exchange angular momentum between the components and a relatively small fraction may be lost from the system due to winds. 
The amount of AM lost is propotional to the mass lost and for stars in this mass range it is small, 
but it can be very large for more massive stars. As the primary
reaches contact, mass transfer starts and a common envelope (CE) forms, hydrodynamic processes 
complete the redistribution of angular momentum in the merged object. We do not know for certain the natal spin angular momentum of
the primary, but there is some evidence that young massive stars are fast rotators with equatorial velocities $v_{\rm eq}\sim$~200~km~s$^{-1}$ 
\citep{rotation,2011ApJ...743L..22D,2013A&A...560A..29R}. 

In the absence of tidal synchronization torques, the spin period at any stage could be estimated from spin angular momentum conservation, 
but more realistically the spin period should be determined taking into acount tidal interactions. 
In order to better understand the process, it is useful to estimate and compare the various timescales involved. 
The relevant timescales to be considered are: the evolutionary times required for the primary
to expand to contact at the above lobe radii and binary separations, the orbital period of the binary at that point, the tidal synchronization 
timescale for the primary to lock to the orbital period, the timescale for the formation of the CE engulfing the secondary, 
and finally the timescale for inspiral and merger. 
 
The evolutionary timescales are calculated by following in {\it MESA} 
\citep{2011ApJS..192....3P,2013ApJS..208....4P,2015ApJS..220...15P,2018ApJS..234...34P,2019ApJS..243...10P} 
the evolution of a single massive star with a mass of either 15~$M_\odot$  or 20~$M_\odot$ from the ZAMS to the 4 expansion stages selected above. 
As is well known, the evolution through the hydrogen burning and the initial expansion to $R_{\rm 1} =$~12~$R_\odot$ is the longest phase, 
taking about $\sim 10^{7}$~yr to complete for the 15~$M_\odot$ primary. The expansion through the gap gradually speeds up, taking 
$\sim 1.2 \times 10^{6}$~yr to reach $R_{\rm 1} =$~100~$R_\odot$, an extra  couple of thousand years to reach
$R_{\rm 1} =$~300~$R_\odot$, and less than a thousand years to expand further to $R_{\rm 1} =$~500~$R_\odot$. These times are of the same order of magnitude 
but somewhat faster and the expansion behavior is qualitatively similar for a more massive primary of 20~$M_\odot$.
 
The orbital period corresponding to first contact is
simply calculated from Kepler's 3rd law taking into account that the primary comes into contact when $R_{\rm 1} =$~0.609$a$ for a mass ratio $q =$~0.067
corresponding to $\mathcal{M}_{\rm 1} =$~15~$M_\odot$, where $a$ is the binary separation \citep{1983ApJ...268..368E}.
The orbital period at the beginning of the plunge phase, when hydrodynamical drag on the companion becomes important, 
can be estimated by setting the binary separation to $R_{\rm 1}$. For the assumed value of $q$, this period is shorter than the period at first contact by a factor 2.1.
The orbital periods for the four assumed fiducial situations are thus $P_{\rm orb} =$~2.5~d, 61~d, 319~d, and 685~d respectively, and slightly shorter for the more massive primary.
 
We estimate the synchronization timescale using Equation 4.12 from the paper by \citet{1977A&A....57..383Z} which is appropriate for stars with a convective 
envelope (that continues to hold for our models out to radii of 300~$R_{\odot}$), and assumes efficient turbulent eddy viscosity. 
\begin{equation}
t_{\rm sync} \sim \frac{1}{6q^{2} k_{\rm 2}}\left(\frac{\mathcal{M}_{\rm 1} R_{\rm 1}^{2}}{L}\right)^{1/3} \frac{I_{\rm 1}}{\mathcal{M}_{\rm 1} R_{\rm 1}^{2}}\left(\frac{a}{R_{\rm 1}}\right)^{6} \label{Eq:sync}
\end{equation}
where $q=\mathcal{M}_{\rm 2}/\mathcal{M}_{\rm 1}$ is the binary mass ratio, $k_{\rm 2}$ is the second order apsidal constant for the primary, $I_{\rm 1}$ is the moment of inertia of the primary and $L$ its luminosity. 
It so happens that $k_{\rm 2}$ and $I_{\rm 1}/\mathcal{M}_{\rm 1} R_{\rm 1}^{2}$ nearly cancel each other to within a factor of two, and given the uncertainties involved, it is sufficient for our purposes to take that ratio as unity. 
Thus the main dependence on parameters comes from the factor $R_{\rm 1}^{2/3}$. These estimates yield $\sim$~100~yr, $\sim$~420~yr, $\sim$~879~yr, 
and $\sim$~1200~yr respectively for spin--orbit synchronization for the cases considered.
 
Upon contact, mass transfer from the primary to the secondary is dynamically unstable and results in 
the formation of a CE leading quickly to engulfing the
secondary into the envelope of the primary. Our 3D hydrodynamic simulation of a 16~$M_{\odot}$+1~$M_{\odot}$ merger at a time when the primary
has expanded out to a radius of $\sim$~12~$R_{\odot}$ (Section~\ref{OctoTiger3D}) show that it only takes $\sim$~5~days for the secondary 
to inspiral and be disrupted as it approaches the core of the primary.
 
A simple analytic estimate for the  inspiral time is obtained by taking the orbital angular momentum of the secondary 
and dividing it by the torque (see Section~\ref{ICs} for details). The angular momentum of the secondary is  $J_{\rm 2}= \mathcal{M}_{\rm 2} a_{\rm 2}^{2} \Omega$, 
where $\Omega$ is the Keplerian orbital frequency, and $a_{\rm 2} = a/(1+q)$ is the distance from the center of mass of the system and the center of mass of the secondary. 
For the hydrodynamic torque we adopt equation (\ref{Eq:drag}), but for the purposes of obtaining and order of magnitude estimate, we set 
$\rho_{\rm 1}(r)$ to the average density of the primary and $\mathcal{M}_{\rm 1}(r) =\mathcal{M}_{\rm 1}$. Therefore we expect our estimate to be a lower limit to the actual inspiral time.
With these approximations we get the following expression for the inspiral time
\begin{equation}
t_{\rm insp} = \frac{8 q^{3/2}}{3 c_{\rm D} (1+q)^{3/2}}\left(\frac{R_{\rm 1}}{R_{\rm 2}}\right)^{7/2}\left(\frac{R_{\rm 2}^{3}}{G \mathcal{M}_{\rm 2}}\right)^{1/2} \label{Eq:insp}
\end{equation}
Inserting the values appropriate for the four scenaria we are contemplating, the above equation yields 4.6~d, 21~yr, 980~yr, and 5900~yr for a primary mass of 
15~$M_\odot$ and correspondingly shorter times, by a factor of 0.65, for a 20~$M_\odot$ primary.
As a check for this approach, we may compare the inspiral timescale obtained in the hydrodynamic simulation for the case $R_{\rm 1} =$~12~$R_\odot$ with our simple analytic estimate. 
 
In summary, with the possible exception of contact at 500~$R_\odot$ and beyond, the initial mass transfer from primary to secondary is likely to start with 
the components synchronous with the orbit. However, as a result of the unstable mass transfer and relatively rapid orbital shrinking to the plunge and 
drag phases, synchronicity may be broken. The inspiral phase is also shorter than, or at most of the same order, as the synchronization time, therefore 
to follow all these processes self--consistently one would have to carry out numerical simulations with all the physics included which is not feasible with the
resources at our disposal. We will therefore adopt the approximation that the secondary spirals inward along a sequence of Keplerian orbits depositing 
a fraction of order unity of the orbital angular momentum in the atmosphere of the primary.
 
\subsection{{\it Merger Model Assumptions}}\label{ICs}

The merger scenario described here is similar to one 
considered by \citep{2002MNRAS.334..819I,2017MNRAS.469.4649M} involving the complete disruption of the secondary
deep inside the envelope of a massive primary during a short--lived CE phase. Such an event is classified as
a ``moderate" merger that eventually leads to evolution toward the red supergiant (RSG) phase. In our model, 
the in--spiral of the secondary is caused solely by the effects of viscosity (dynamical friction, or drag) of the secondary within the envelope
of the primary. During this process a fraction of the angular momentum lost by the secondary is deposited to the envelope of the
primary causing it to spin up. The dynamical merger process begins at the time of contact between the expanding envelope of the primary and 
the secondary and continues until the secondary reaches a radius equal to its tidal radius where it gets complete disrupted. 
This effective tidal radius is very close to the radius of the He core of the primary ($\sim$~1~$R_{\odot}$).
A main difference between this work and that of \citep{2017MNRAS.469.4649M} is our predictions 
about the effect of such merger to the evolution of the angular momentum profile of the primary and, more specifically,
the surface equatorial rotational velocity.

To derive an approximate analytical model of the perturbed internal structure of a massive primary star
in the context of the merger scenario discussed above, we resort to the following assumptions for the progenitor binary system:
\begin{enumerate}[label=\roman*.]
\item{A low mass--ratio for the binary progenitor system ($q = \mathcal{M}_{\rm 2}/\mathcal{M}_{\rm 1} =$~0.07--0.25). 
For our study we adopt binary systems with ($\mathcal{M_{\rm 1}}$,$\mathcal{M_{\rm 2}}$)~$=$ (15, 1), (16, 4) and (17, 3) (in solar units)
that satisfy this criterion.
The choice of a low mass--ratio allows for the possibility that the secondary
spends a considerable amount of time in the envelope of the primary while spiraling--in toward the
core, well before it is tidally disrupted. This guarantees that, in the process, the secondary deposits a fraction
of its angular momentum into the envelope of the primary and that the initial RLOF is dynamically unstable and
will result in a CE. Additional arguments about the necessity of a low mass--ratio
were discussed in Section~\ref{merger_necessary}.}
\item{The merger event occurs after the end of the primary's main--sequence phase and
while its envelope is expanding while crossing the ``Hertzsprung gap'' over a thermal time--scale (lasting
a few tens of thousands of years). As such, our models consider early Case B merger events.
This implies that the envelope of the primary comes into contact
with the secondary due to the expansion of the former in its way to becoming a supergiant star and at radii of
200~$<R_{\rm 1}<$~700~$R_{\odot}$, where $R_{\rm 1}$ is the radius of the primary. This condition 
is necessary in order to facilitate a large density contrast between the secondary and the envelope of the
primary ensuring that the secondary remains intact during the coalescence.}
\item{The spiral--in time--scale of the secondary is significantly smaller than the thermal adjustment
time--scale of the envelope of the primary. This approximation implies that in our model we do not
take into account the effect of the expansion of the primary's envelope during the merger event and assume
that it does not affect the dynamics of the spiral--in process. This is a good approximation given that, for the above--mentioned
initial binary conditions during contact, the time--scale it takes for the secondary to reach the tidal radius (which is approximately
equal to the radius of the helium (He) core of the primary) is in the order of $\sim$~100~years 
while the thermal--adjustment time--scale of the primary's envelope is in the order of $\sim$~10$^{4}$~years. }
\item{The magnitude of the secondary's velocity during the spiral--in is assumed to be equal to the corresponding Keplerian value at all times:
$v_{\rm 2} = \sqrt{G \mathcal{M}_{\rm 1} (r) / r}$ where $v_{\rm 2}$ is the velocity magnitude of the secondary, $G$ the gravitational constant and
$\mathcal{M}_{\rm 1}(r)$ the mass of the primary enclosed within radial coordinate $r$.}
\item{Changes in the internal structure of the secondary are ignored during the merger event and it is approximated
as a point source. This removes the complication of considering effects such as tidal deformation of the secondary and tidal friction
on the dynamical evolution of the merger. For more details on these effects three--dimensional hydrodynamic simulations
are necessary \citep{1990A&A...227L...9P,2002PhDT........25I,2002MNRAS.334..819I}.}
\item{The effects of the merger on nucleosynthesis are ignored. The merger hypotheses discussed in this work imply the mixing of
1--4~$M_{\odot}$ of material from a main--sequence secondary star in the hydrogen (H) envelope and the He core
of the primary potentially leading to distinct nucleosynthetic signatures. Following that distinct first, dynamical phase of merger--induced mixing, in the longer term and
depending on the convective and rotational mixing time--scales in the envelope of the primary, it is possible to dredge--up enhanced
nucleosynthetic products of the CNO process to the photosphere of the post--merger. In addition, the H--fuel deposited from the secondary will increase the mass of the H--burning
shell around the He--core of the primary resulting in higher luminosity which may cause the primary envelope to expand to a larger radius. We aim
to study the nucleosynthetic signatures of such merger events in a future study (see also \citealt{2002MNRAS.334..819I}).}
\item{In general we expect the secondary to enter the envelope of the primary at supersonic speeds generating shocks 
in both the ambient medium and the secondary star's atmosphere. A complex flow will be set up in which the shocked primary material 
and some of the secondary atmosphere will be mixed and entrained in the turbulent wake behind the secondary. 
Consequently there will be an increase in the specific entropy of the perturbed layers, but the complexity of the flow prevents us 
from calculating this increase in detail. We adopt an approximate approach that shows that the increase can be neglected. 
The change in specific entropy in a planar shock is given by 
\begin{equation}
\Delta s = c_{\rm V} \left[\ln{(p_{\rm 2}/p_{\rm 1})} - \gamma\ln{(\rho_{\rm 2}/\rho_{\rm 1})}\right],\label{Eq:Ds}
\end{equation}
where the suffixes 1 and 2 indicate pre-- and post--shock values respectively. The pressure and density ratios can be calculated 
given the pre--shock Mach number and the adiabatic index $\gamma$ \citep{1959flme.book.....L}. 
Consistent with our assumptions we take the Mach number to be given by the ratio of the local Keplerian velocity to the local sound speed, and 
using values appropriate for the envelope of the primary we can calculate the change in specific entropy. This turns out to be 
typically at the level of 1\% compared to the local entropy of the primary throughout most of the star and only becomes comparable near the surface along
the stellar equator. However, this entropy increase is limited to a toroidal region of cross section $\sim R_{\rm 2}^2$ in the wake of the secondary, 
whereas {\it MESA}, being a 1D stellar evolution code, expects the entropy increase to be distributed in a shell of radius $r$ and thickness $R_{\rm 2}$, so the entropy increase is 
further diluted by a factor $\sim R_{\rm 2}/r$, and thus the entropy increase due to the inspiral is small compared to the ambient entropy of the primary everywhere. 
In conclusion,  we take the post--merger entropy profile of the primary to be, in good approximation, the same as the entropy profile right before the merger.}
\end{enumerate}

Given those simplifying assumptions, our approach aims to be a simple proof--of--principle effort to illustrate the potential of merger events
to affect the rotational properties of massive stars past the main--sequence and an effort to reproduce the observed properties of
Betelgeuse under the assumption that it suffered such a merger.

\subsection{{\it Derivation of Post--Merger Structure}}\label{perturbed_prof}

In order to derive an expression for the perturbed internal distribution of specific angular momentum in the primary star following
a merger event with a smaller companion, we first consider the action of a torque that acts on the secondary upon
entering the evenlope of the primary. This torque originates due to dynamical friction (drag) and has the generic form:
\begin{equation}
\tau(r) = -\frac{1}{2} c_{\rm D} \pi R_{2}^{2} \rho_{\rm 1}(r) v_{\rm 2}^{2} r,\label{Eq:drag}
\end{equation}
where $c_{\rm D}$ is the drag coefficient, $R_{2}$ the radius of the secondary, $\rho_{\rm 1}(r)$ the density profile of the primary at radial coordinate
$r$ and $v_{\rm 2}$ the magnitude of the secondary's velocity which we are assuming to be equal to the Keplerian value at all times (see Section~\ref{ICs}).
The basic dynamic equation is then:
\begin{equation}
\mathcal{M}_{\rm 2} \frac{d}{dt}\sqrt{G\mathcal{M}_{\rm 1}(r) r} = -\frac{1}{2} c_{\rm D} \pi R_{2}^{2} \rho_{\rm 1}(r) G \mathcal{M}_{\rm 1}(r),\label{Eq:motion}
\end{equation}
where $\mathcal{M_{\rm 2}}$ is the mass of the secondary. Numerical solutions of Equation~\ref{Eq:motion},
which is a simple ordinary differential equation yield the evolution of the radial coordinate of the secondary, $r(t)$.

Over an infinitesimal radial displacement equal to $dr$, the angular momentum of the primary, $J_{\rm 1}$, increases by an amount equal to that lost due to
the spiral--in of the secondary, $J_{\rm 2}$ as follows:  
\begin{equation}
\frac{dJ_{\rm 1}}{dr} = f \frac{1}{\dot{r}}\frac{dJ_{\rm 2}}{dt}.\label{Eq:angmom}
\end{equation}
The factor $f\le 1$ is the fraction of the angular momentum lost by the secondary that results in bulk rotation of the layers of
the primary, the rest appearing as local vorticity and being eventually dissipated.
Accordingly, the {\it specific} angular momentum deposited in the primary is equal to:
\begin{equation}
j_{\rm spin} (r) = \frac{dJ_{\rm 1}/dr}{\frac{d\mathcal{M}_{\rm 1}}{dr}} = f \frac{d J_2}{d\mathcal{M}_{\rm 1}}.\label{Eq:js1}
\end{equation}
%$f$ is an efficiency factor taken to be of the order of unity in our analysis and 

% By analyzing the left hand side of Equation~\ref{Eq:motion} and expanding the derivative $dJ_{\rm 2}/dt = d(\mathcal{M}_{\rm 2} v_{\rm 2} r)/dr$ we obtain
Taking $J_2 = \mathcal{M}_{\rm 2} \sqrt{G \mathcal{M}_{\rm 1} (r) r}$, with $\mathcal{M}_{\rm 2}$ constant, we obtain
the deposited specific angular momentum profile of the primary following the merger event for $R_{\rm 1,He}<r_{\rm 2}<R_{\rm 1}$, 
where $R_{\rm 1,He}$, $r_{\rm 2}$ and $R_{\rm 1}$ is the radius of the He--core of the primary, the radial distance of the secondary from
the center of the primary and the radius of the primary respectively:
\begin{equation}
j_{\rm spin} (r) = 
%\frac{\mathcal{M}_{\rm 2} \sqrt{G \mathcal{M}_{\rm 1} (r)}}{8 \pi f} \left[ \frac{1}{r^{5/2} \rho_{\rm 1}(r)}+\frac{8 \pi r^{1/2}}{\mathcal{M}_{\rm 1} (r)} \right].
\frac{f q}{2}\sqrt{G \mathcal{M}_{\rm 1} (r) r}\left(1+\frac{d\ln{r}}{d\ln{\mathcal{M}_{\rm 1}}}\right),
\label{Eq:js2}
\end{equation}
where $d\mathcal{M}_{\rm 1}/dr=4 \pi r^{2} \rho(r)$.
This expression shows that the angular momentum per unit mass deposited is proportional to the specific angular momentum of the Keplerian orbit of the secondary
times a function that rises steeply toward the surface of the primary. This can be seen clearly by considering an alternative expression for the logarithmic derivative 
above in terms of the ratio between the average density of the primary inside radius $r$ and the local density at that radius 
($d\ln{r}/d\ln{\mathcal{M_{\rm 1}}} = \overline{\rho_{\rm 1}}(r)/3 \rho_{\rm 1}(r)$).
We assume that the secondary completely dissolves when $r \simeq$~$R_{\rm 1, He}$ due to strong tidal forces, so that $j_{\rm spin} (r) = 0$ for $r<R_{\rm 1, He}$. 
If $j_{\rm 1} (r)$ is the original, un--perturned specific angular momentum profile of the primary then, following the merger event, the final perturbed profile will be:
\begin{equation}
\widehat{j_{\rm 1}} (r) = j_{\rm 1} (r) + j_{\rm spin} (r).\label{Eq:js3}
\end{equation}

\begin{figure*}
\gridline{\fig{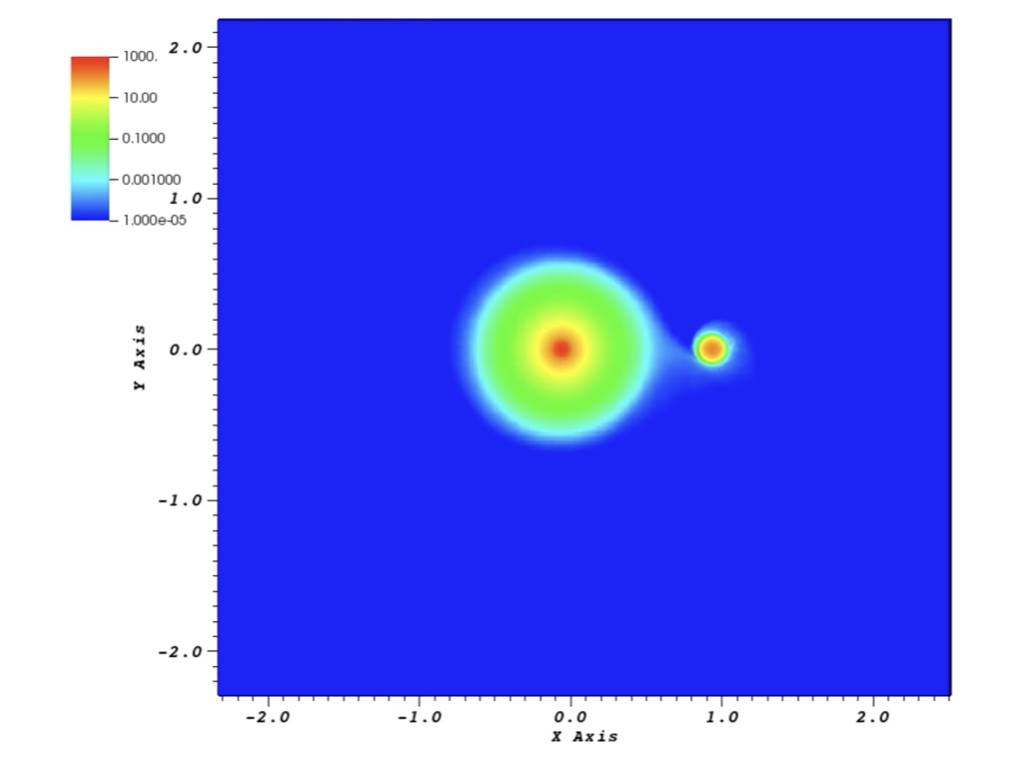}{0.5\textwidth}{}
          \fig{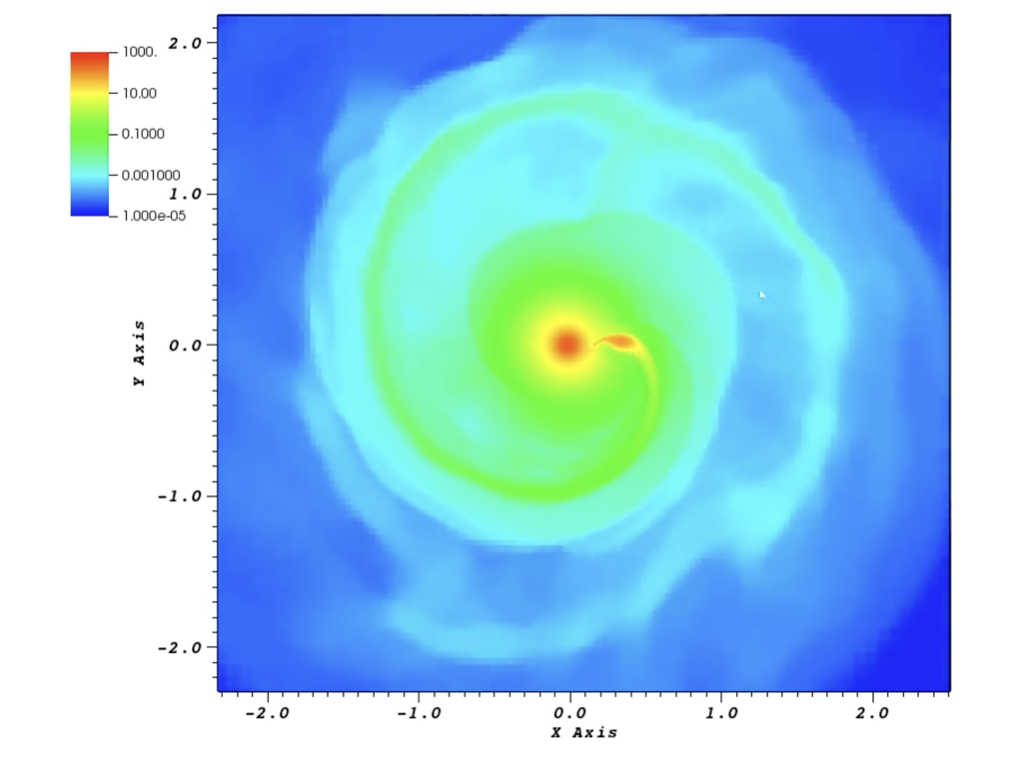}{0.5\textwidth}{}
          }
\caption{Face--on view of the {\it OctoTiger} 16~$M_{\odot}$+1~$M_{\odot}$ merger simulation occurring at $R_{\rm 1} =$~12~$R_{\odot}$ during the beggining of
the simulation ({\it left panel}) and the tidal--disruption of the secondary around the core of the primary ({\it right panel}). Different colors correspond to different
values for gas density as shown in the colorbars. The X and Y axis scales are in units of the initial binary separation of the binary (12~$R_{\odot}$).
\label{Fig:mergersim}}
\end{figure*}

\subsection{{\it 3D Simulation of the $R_{\rm 1} =$~12~$R_{\odot}$ merger case.}}\label{OctoTiger3D}

In order to numerically evaluate the properties of an early Case B merger and derive
a better measure of the inspiral time (Section~\ref{assump}), we run a 3D merger simulation for a 16~$M_{\odot}$+1~$M_{\odot}$ system 
with the parallelized adaptive mesh refinement (AMR) code {\it OctoTiger} \citep{2007ApJ...670.1314M,2016MNRAS.462.2237K}. 
Due to computational limitations, were only able to properly resolve
the $R_{\rm 1} =$~12~$R_{\odot}$ case in 3D. At this phase, the primary is just past its TAMS with
no H burning taking place in the core and it is starting its Hertzsprung gap--crossing phase. 
For the larger primary radius choices (100, 300, 500~$R_{\odot}$), the density contrast between the secondary 
and the envelope of the primary would be such that it would require a large degree of mesh refinement and a prohibitive number of computing cells. 
We argue, however, that the final phase of this simulation corresponding to the plunge of the secondary into the central regions of the primary eventually leading
to its tidal disruption, will not be singificantly different for the other cases.

To initialize the 3D cartesian grid of {\it OctoTiger}, we first compute the structures of both the 16~$M_{\odot}$ at 12~$R_{\odot}$
and the 1~$M_{\odot}$ components using variations of the publicly available {\tt 15M\_dynamo} and 
{\tt 1M\_pre\_ms\_to\_wd} {\it MESA} test suite problems accordingly. 
Our {\it MESA} input inlist files for these runs and all of our merger models will be made publicly available at the 
{\it MESA marketplace}\footnote{www.mesastar.org} website.
We then fit the structure of each of the components of the binary system
with bipolytropic functions as required for the initialization method used in {\it OctoTiger} \citep{2016MNRAS.462.2237K}. 
For the 16~$M_{\odot}$ primary during the phase when its radius is 12~$R_{\odot}$ the bipolytropic indices were found to be 3.2 for the core and 3.1 for the envelope. 
For the 1~$M_{\odot}$ secondary, accordingly, the best--fit bipolytripc indices were 3.0 and 1.5 for the core and the envelope respectively.
Upon mapping to {\it OctoTiger} the dynamical merger was driven by increasing the entropy of the primary's envelope \citep{2016MNRAS.462.2237K}. 
Within $\sim$~175 orbits the secondary came into contact with the primary and subsequently experienced an inspiral toward the He core of the primary until it got
tidally disrupted. This latter phase lasted $\sim$~5 days, and is in excellent agreement with the simple analytical estimate for the inspiral timescale
that we provided in Section~\ref{assump}.
Figure~\ref{Fig:mergersim} shows a face--on view of the initial binary arrangement (left panel) and the final phase of secondary tidal disruption (right panel) 
as simulated in {\it OctoTiger}. 

\section{Stellar Evolution Calculations}\label{evol}

\begin{figure*}
\begin{center}
\includegraphics[angle=0,width=18cm,trim=0.in 0.25in 0.5in 0.15in,clip]{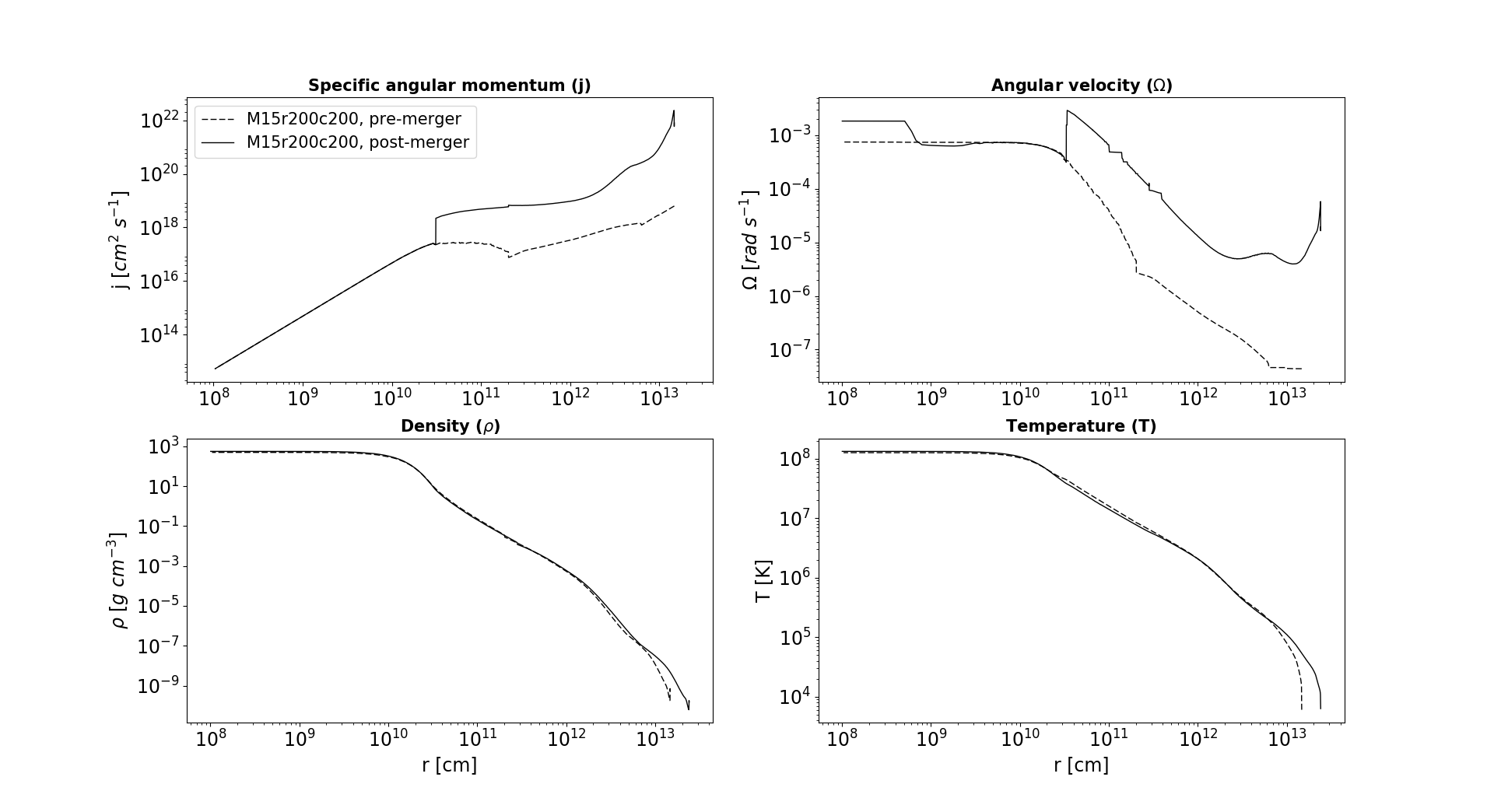}
\caption{Comparison of pre-- and post--merger internal distribution profiles for 
specific angular momentum $j$ ({\it upper left panel}),
angular velocity $\Omega$ ({\it upper right panel}),
density $\rho$ ({\it lower left panel}) and
temperature $T$ ({\it lower right panel}) of the 15~$M_{\odot}$ primary rotating at 200~km~s$^{-1}$ during the ZAMS
and for a contact radius of 200~$R_{\odot}$ (model ``M15r200c200'').}
\label{Fig:15sm_mapping}
\end{center}
\end{figure*}

%\begin{figure*}
%\begin{center}
%\includegraphics[angle=0,width=18cm,trim=0.in 0.25in 0.5in 0.15in,clip]{20Msun_300Rsun_mapping.png}
%\caption{Same as in Figure~\ref{Fig:15sm_mapping} but for model ``M20r200c300''.}
%\label{Fig:20sm_mapping}
%\end{center}
%\end{figure*}

\setcounter{table}{0}
\begin{deluxetable*}{lccccccccc}
\tabletypesize{\footnotesize}
\tablecaption{Properties of the {\it MESA} models considered in this work.}
\tablehead{
\colhead {Model} &
\colhead{$\mathcal{M}_{\rm 1}$~($M_{\odot}$)} &
\colhead{$\mathcal{M}_{\rm 2}$~($M_{\odot}$)} &
\colhead {$v_{\rm 1,rot,i}$~(km~s$^{-1}$)} &
\colhead{$R_{\rm 1}$~($R_{\odot}$)} &
\colhead {$\log T_{\rm eff, 0}$~(K)} &
\colhead {$\log(L_{\rm 0}/L_{\odot})$} &
\colhead {$v_{\rm rot, 0}$~(km~s$^{-1}$)} &
\colhead {$\Delta t_{\rm target}$~(yr)} &
\\}
\startdata
 & & & &{\it Single star evolution}& & & & \\
\hline
\hline
& & & & & & & & \\
S15r0 & 15 & -- & 0 &  -- & 3.50 & 5.02 & 0.0 & 0.0  \\
S15r200 & 15 & -- &  200 & -- & 3.49 & 5.04 & 0.15 & 35.0 \\
S15r200b$^{\dagger}$ & 15 & -- & 200 & -- & 3.50 & 5.02 & 0.18 & 0.0 \\
S15r300 & 15 & -- & 300 & -- & 3.49 & 5.03 & 0.18 & 66.7 \\
S15r500 & 15 & -- & 500 & -- & 4.51 & 5.62 & 436.23 & 0.0 \\
S20r0 & 20 & -- & 0 & -- & 3.49 & 5.30 & 0.0 & 0.0 \\
S20r0b$^{\ddagger}$ & 20 & -- & 0 & -- & 3.49 & 5.19 & 0.0 & 0.0 \\
S20r200 & 20 & -- & 200 & -- & 3.49 & 5.32 & 0.06 & 0.0 \\
\hline
& & & &{\it Binary merger, $q =$~0.07}& & & & \\
\hline
M15r0c200 & 15 & 1 & 0 & 200 & 3.50 & 5.04 & 5.00 & $8.3 \times 10^{4}$ \\
M15r0c300 & 15 & 1 & 0 & 300 & 3.50 & 5.03 & 5.98 & $2.6 \times 10^{5}$ \\
M15r0c700 & 15 & 1 & 0 & 700 & 3.48 & 5.03 & 20.40 & $1.7 \times 10^{5}$ \\
M15r200c200 & 15 & 1 & 200 & 200 & 3.50 & 5.04 & 4.66 & $1.5 \times 10^{4}$ \\
M15r200c300 & 15 & 1 & 200 & 300 & 3.50 & 5.04 & 5.59 & $1.6 \times 10^{5}$ \\
M15r200c700 & 15 & 1 & 200 & 700 & 3.48 & 5.04 & 19.95 & $1.2 \times 10^{5}$ \\
\hline
& & & &{\it Binary merger, $q =$~0.18}& & & & \\
\hline
M17r0c200 & 17 & 3 & 0 & 200 & 3.50 & 5.24 & 6.83 & $5.0 \times 10^{5}$ \\
M17r0c250 & 17 & 3 & 0 & 250 & 3.50 & 5.24 & 6.81 & $5.1 \times 10^{5}$ \\
M17r0c300 & 17 & 3 & 0 & 300 & 3.50 & 5.24 & 7.49 & $5.3 \times 10^{5}$ \\
M17r0c200\_vL & 17 & 3 & 0 & 200 & 3.50 & 5.24 & 6.82 & $5.0 \times 10^{5}$ \\
M17r0c250\_vL & 17 & 3 & 0 & 250 & 3.50 & 5.24 & 6.79 & $5.1 \times 10^{5}$ \\
M17r0c300\_vL & 17 & 3 & 0 & 300 & 3.50 & 5.23 & 7.48 & $5.3 \times 10^{5}$ \\
M17r200c200 & 17 & 3 & 200 & 200 & 3.50 & 5.29 & 4.35 & $6.9 \times 10^{5}$ \\
M17r200c250 & 17 & 3 & 200 & 250 & 3.50 & 5.30 & 6.18 & $6.9 \times 10^{5}$ \\
M17r0c300 & 17 & 3 & 200 & 300 & 3.50 & 5.31 & 8.94 & $6.8 \times 10^{5}$ \\
\hline
& & & &{\it Binary merger, $q =$~0.25}& & & & \\
\hline
M16r0c200 & 16 & 4 & 0 & 200 & 3.50 & 5.22 & 9.10 & $8.0 \times 10^{5}$ \\
M16r0c250 & 16 & 4 & 0 & 250 & 3.50 & 5.21 & 10.90 & $7.8 \times 10^{5}$ \\
M16r0c300 & 16 & 4 & 0 & 300 & 3.50 & 5.20 & 12.05 & $7.6 \times 10^{5}$ \\
M16r200c200 & 16 & 4 & 200 & 200 & 3.50 & 5.28 & 4.50 & $7.1 \times 10^{5}$ \\
M16r200c250 & 16 & 4 & 200 & 250 & 3.50 & 5.27 & 6.17 & $7.2 \times 10^{5}$ \\
M16r200c300 & 16 & 4 & 200 & 300 & 3.50 & 5.25 & 8.94 & $7.0 \times 10^{5}$ \\
\enddata 
\tablecomments{$^{\dagger}$This model was run without the effects of magnetic fields on angular momentum transport
and chemical mixing turned--on.
$^{\ddagger}$This model was run with enhanced mass--loss during the main--sequence and post--main sequence phase (by selecting
a higher mass--loss coefficient for the Vink and Reimers prescriptions used).
\label{T1}}
\end{deluxetable*}

\begin{figure*}
\gridline{\fig{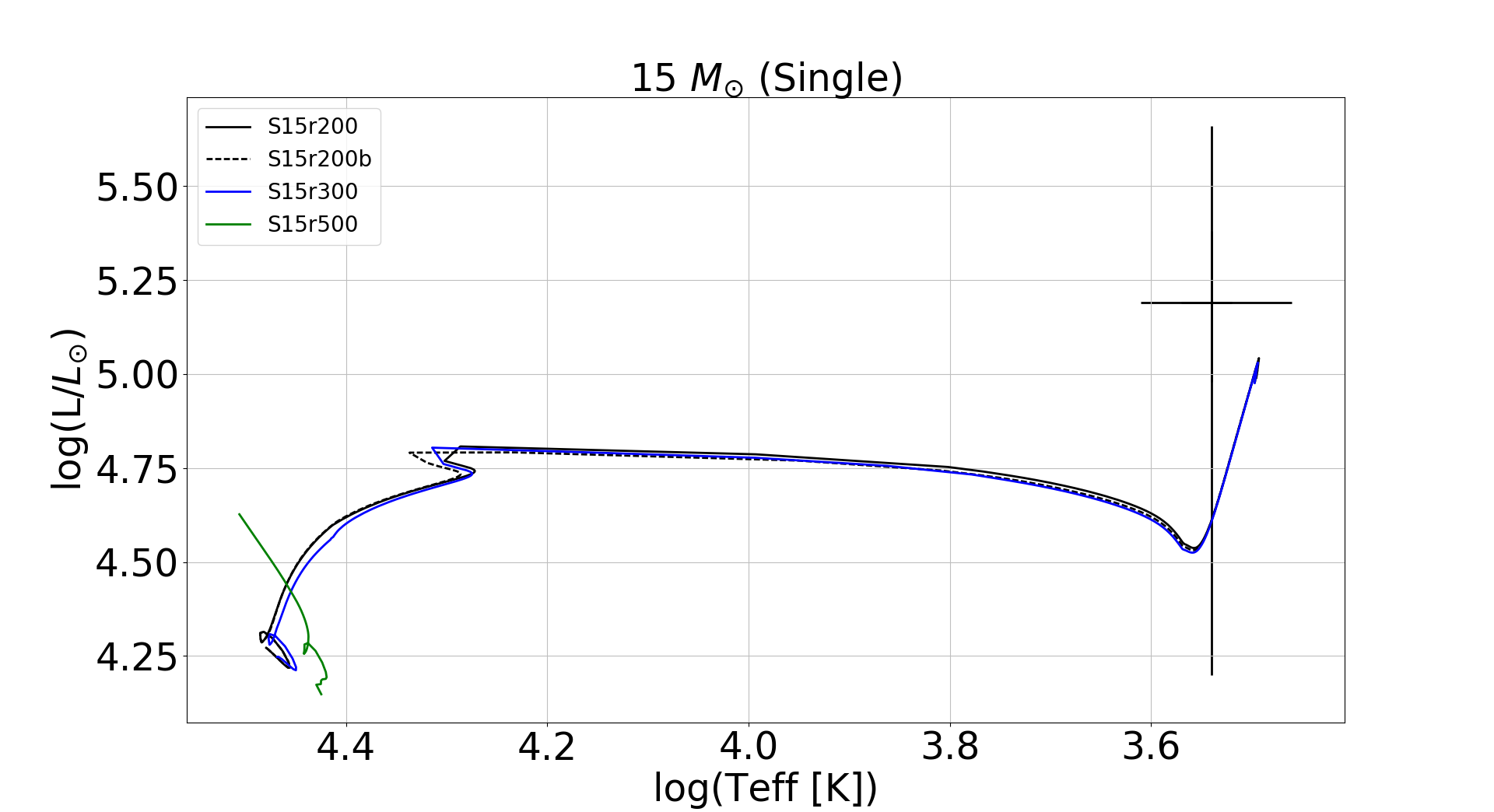}{0.5\textwidth}{}
          \fig{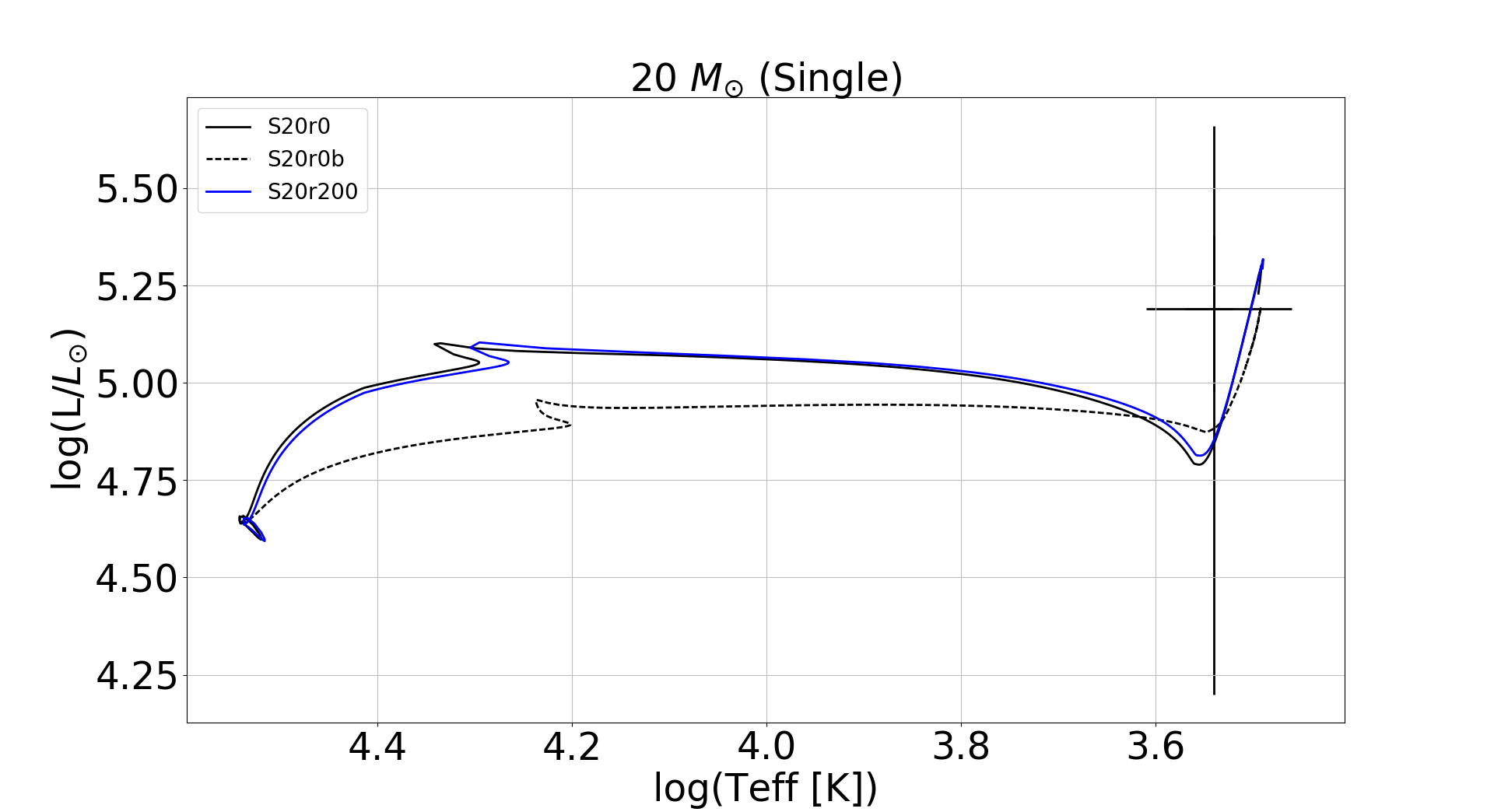}{0.5\textwidth}{}
          }
\caption{Evolution of the non--rotating and rotating 
single (``S'') models in the Hertzsprung–-Russell diagram for ZAMS mass of 
15~$M_{\odot}$ ({\it left panel}) and
20~$M_{\odot}$ ({\it right panel}).
The 3-$\sigma$ observed location of Betelgeuse is also marked for comparison.
\label{fig:single}}
\end{figure*}

\begin{figure*}
\gridline{\fig{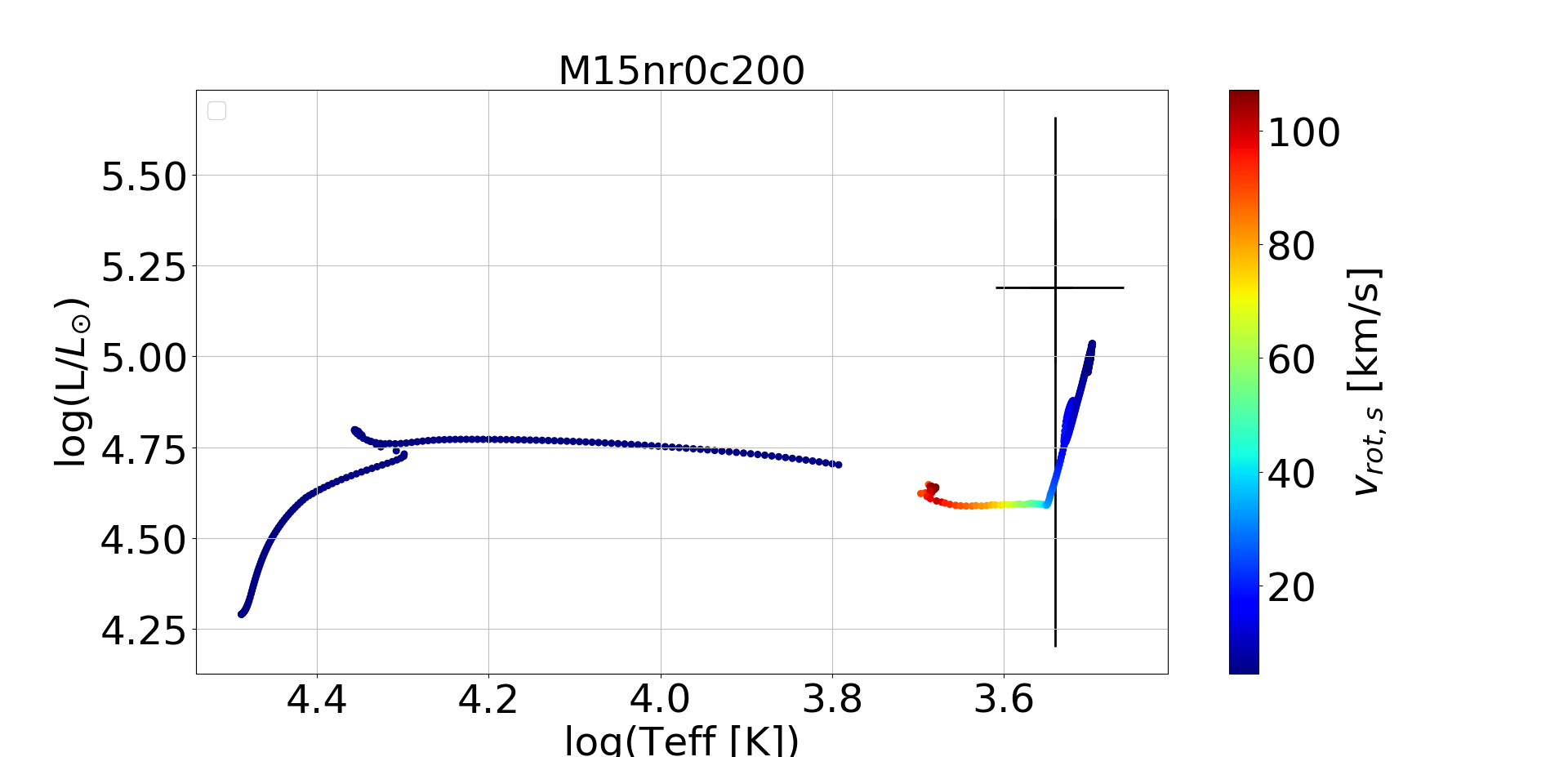}{0.5\textwidth}{}
          \fig{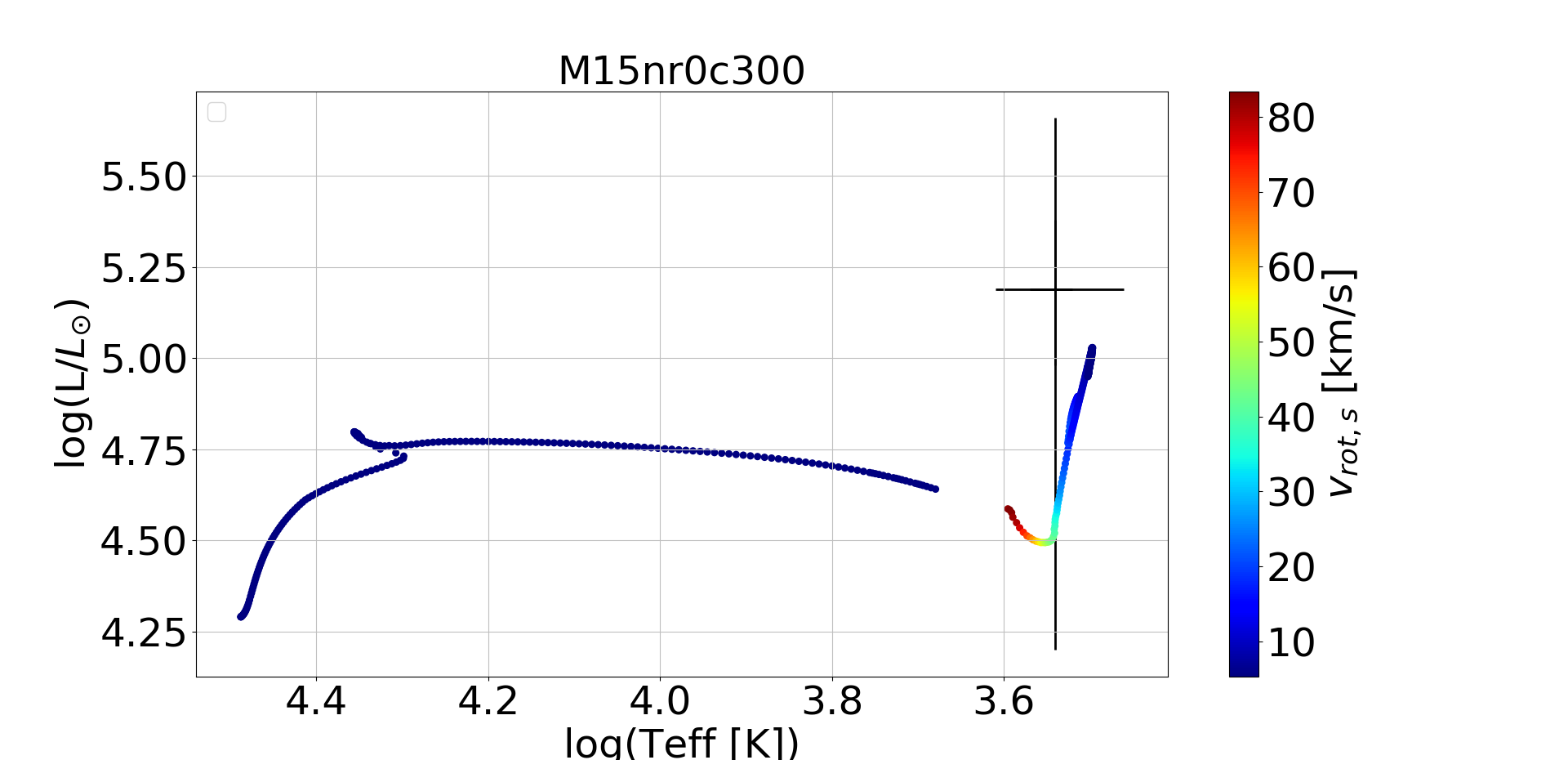}{0.5\textwidth}{}
          }
\gridline{\fig{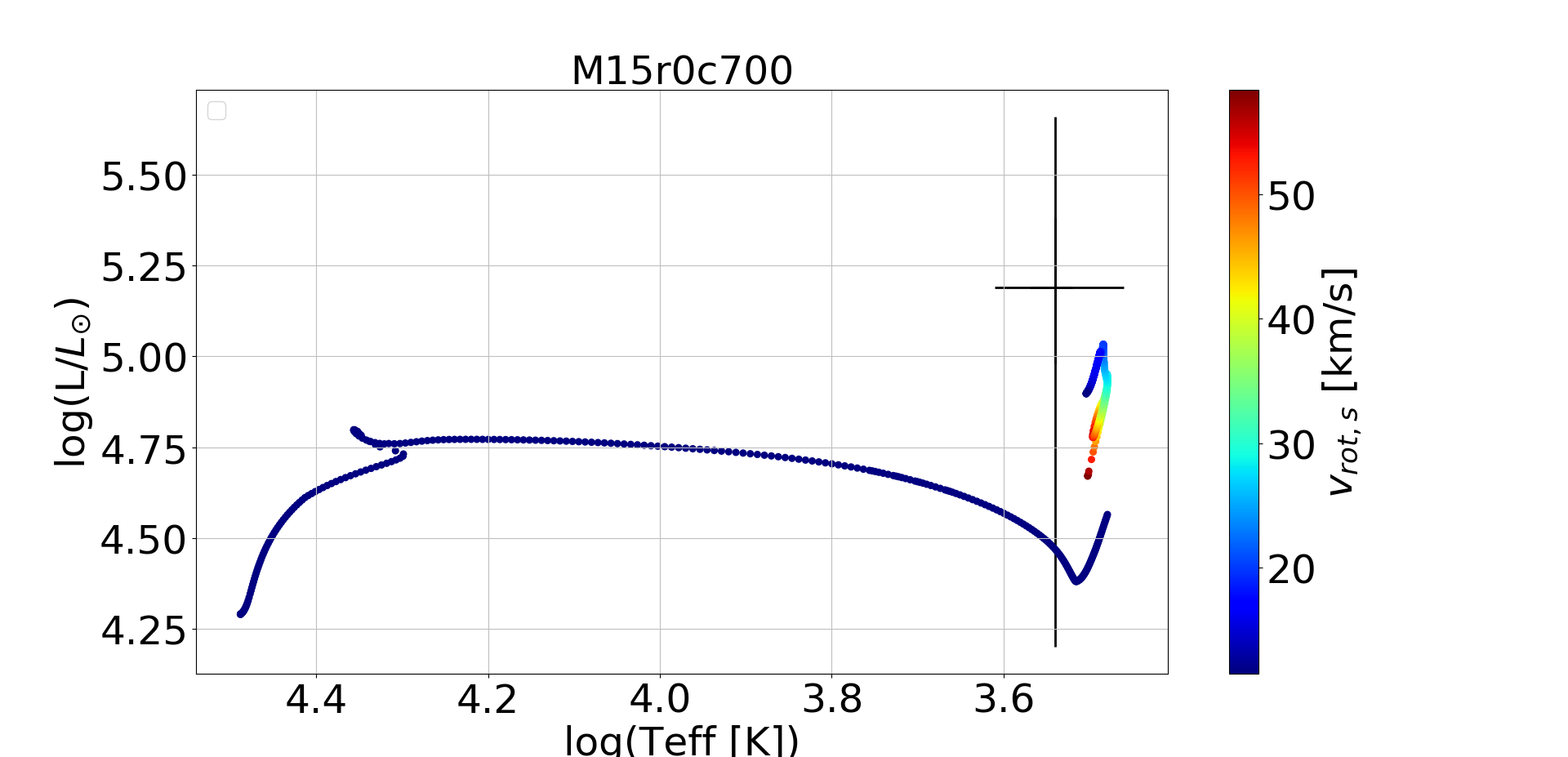}{0.5\textwidth}{}
          \fig{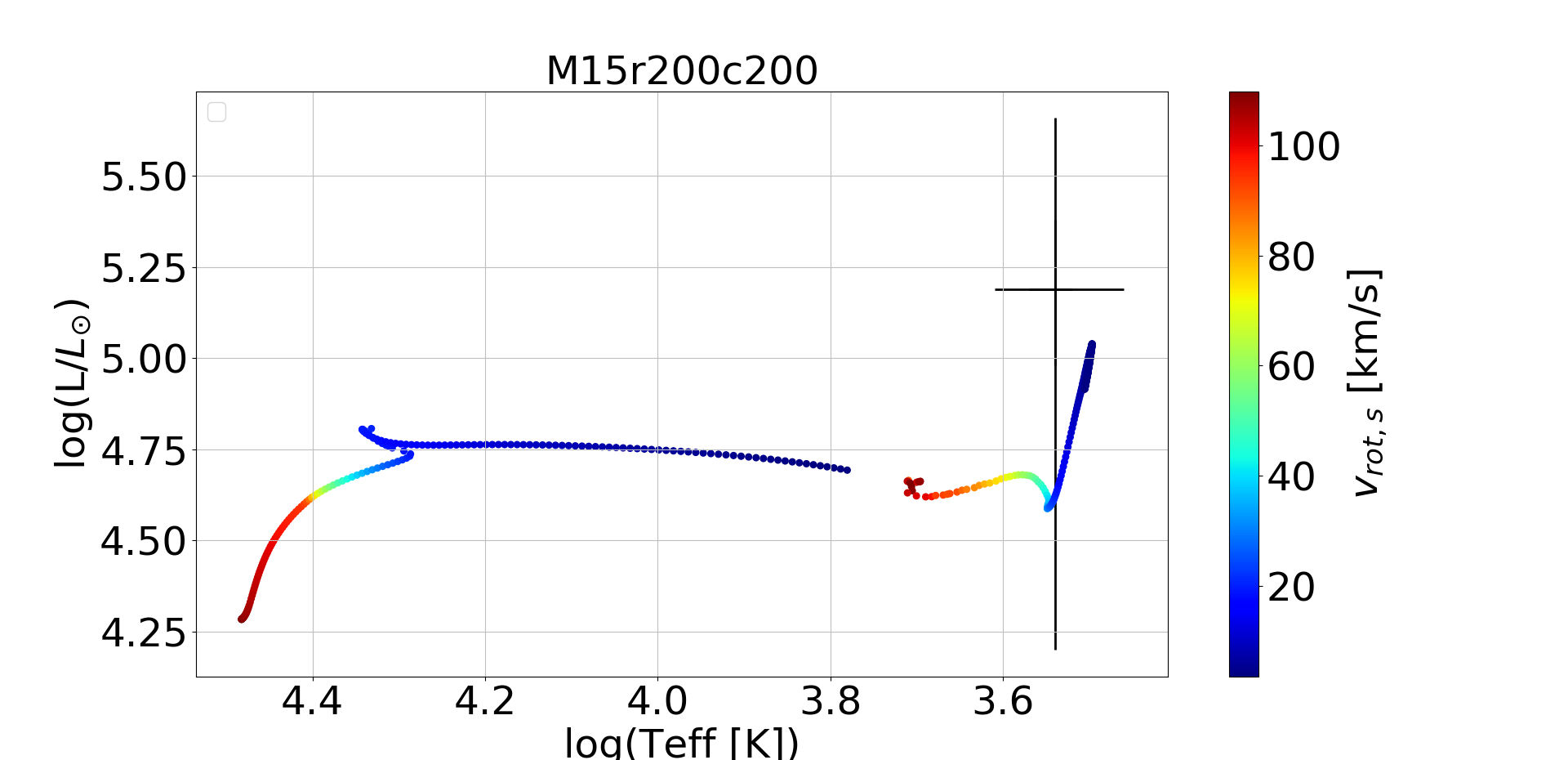}{0.5\textwidth}{}
          }
\gridline{\fig{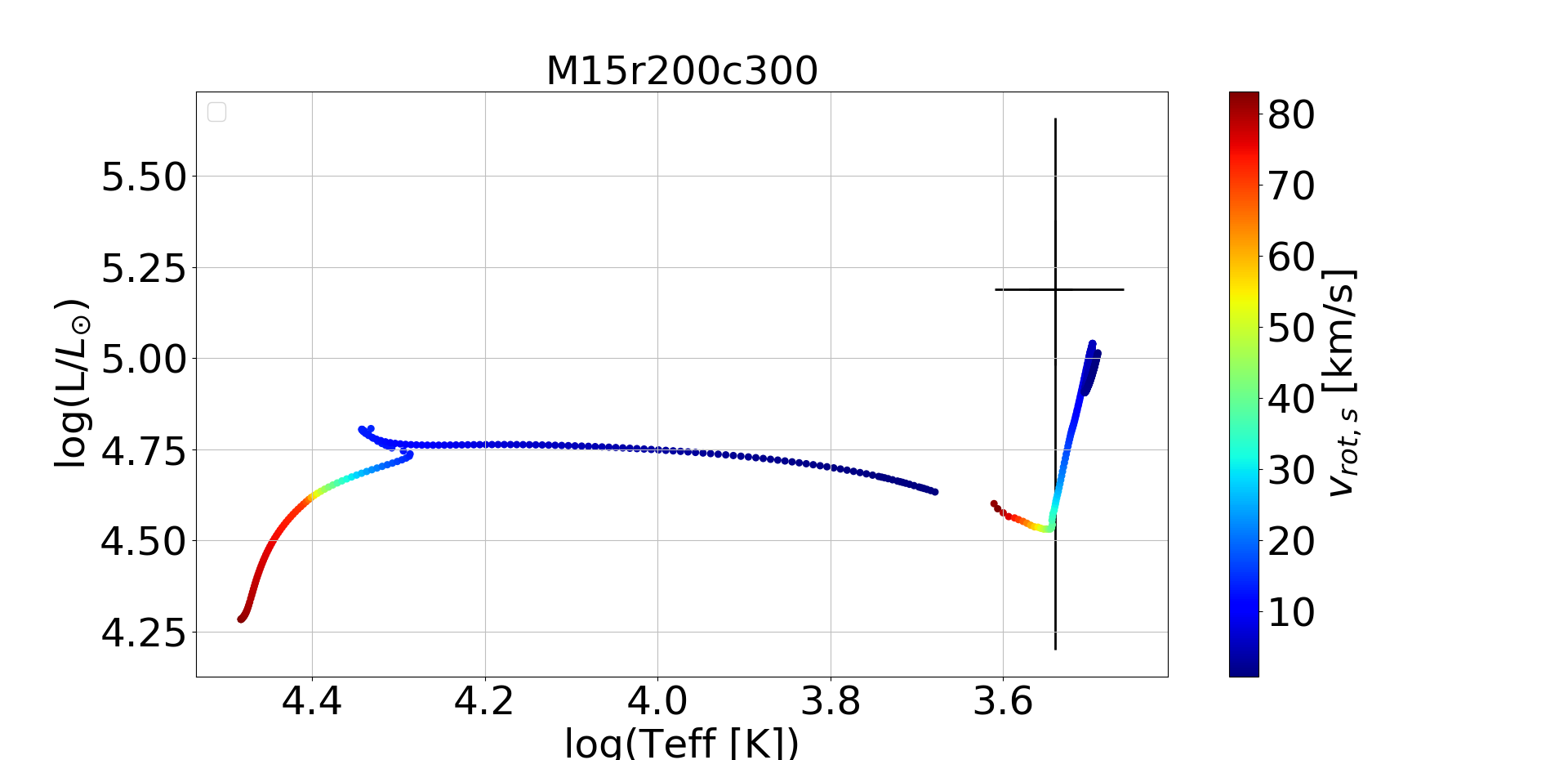}{0.5\textwidth}{}
          \fig{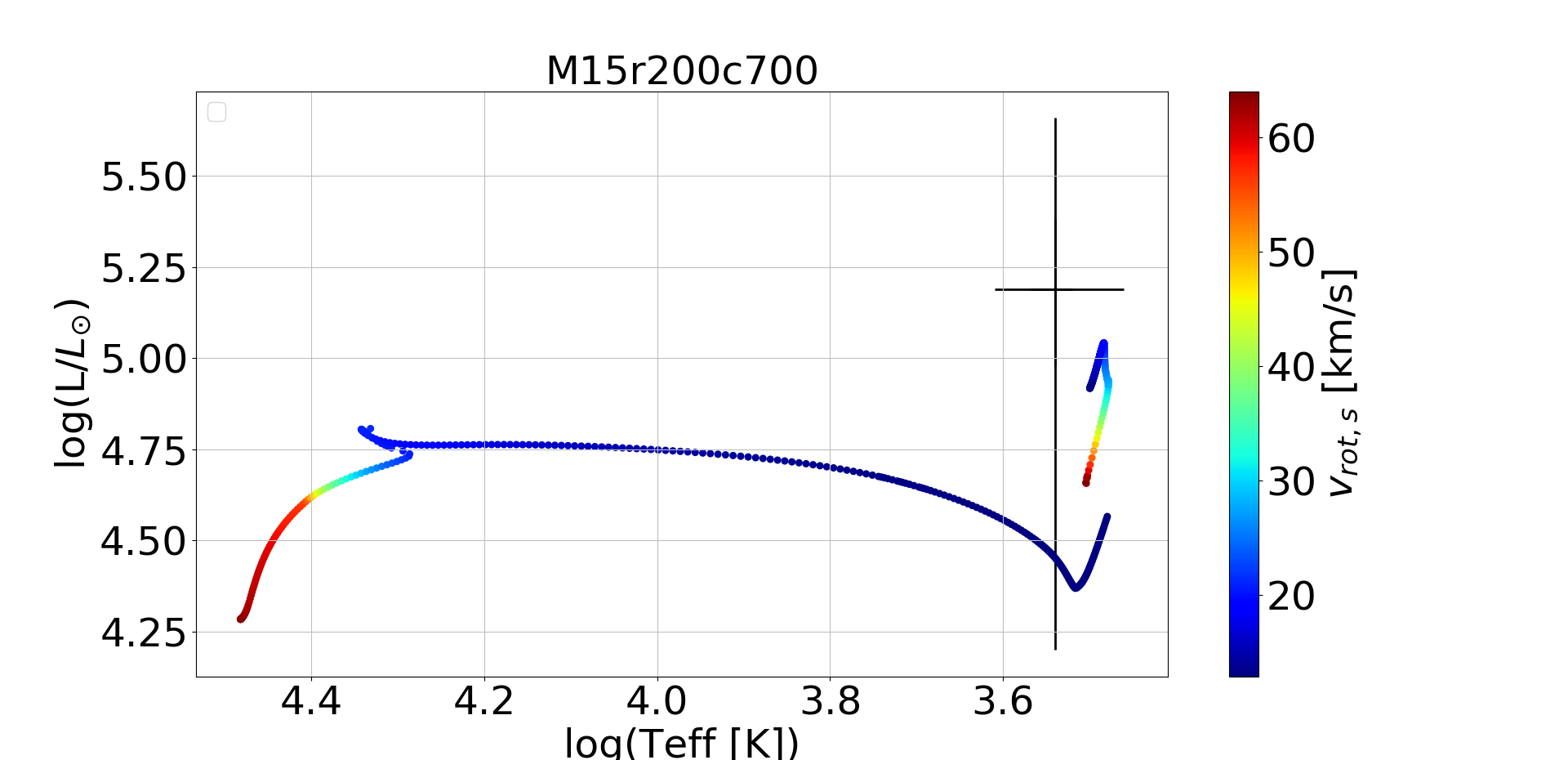}{0.5\textwidth}{}
          }
\caption{Evolution of the non--rotating and rotating 
merger (``M'') models in the Hertzsprung–-Russell diagram for ZAMS primary and secondary mass of 
15~$M_{\odot}$ and 1~$M_{\odot}$ accordingly ($q =$~0.07; {\it left panel}).
The 3-$\sigma$ observed location of Betelgeuse is also marked for comparison.
The colorbar shows the corresponding surface equatorial rotational velocity values
at different evolutionary stages. 
\label{fig:M15}}
\end{figure*}

\begin{figure*}
\gridline{\fig{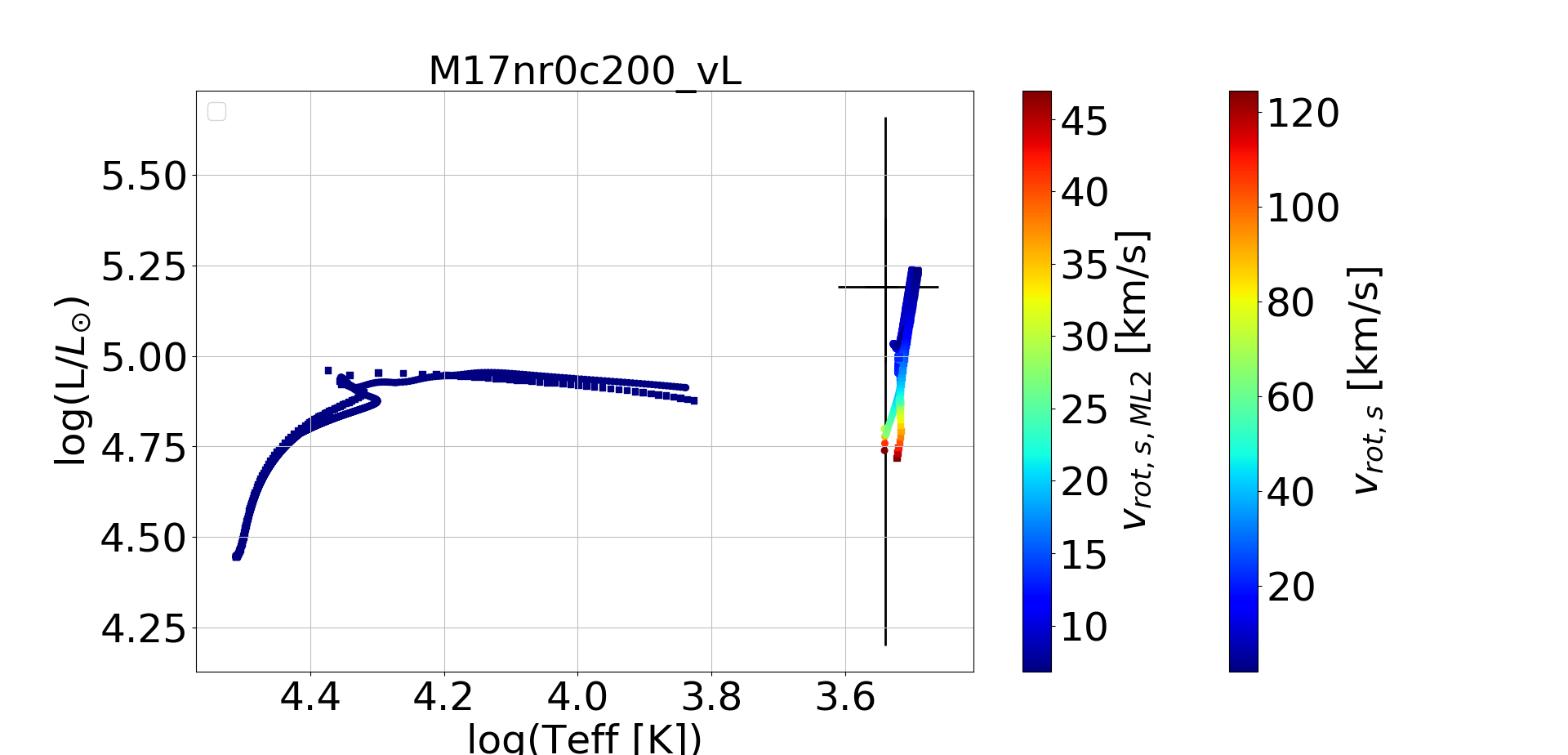}{0.5\textwidth}{}
          \fig{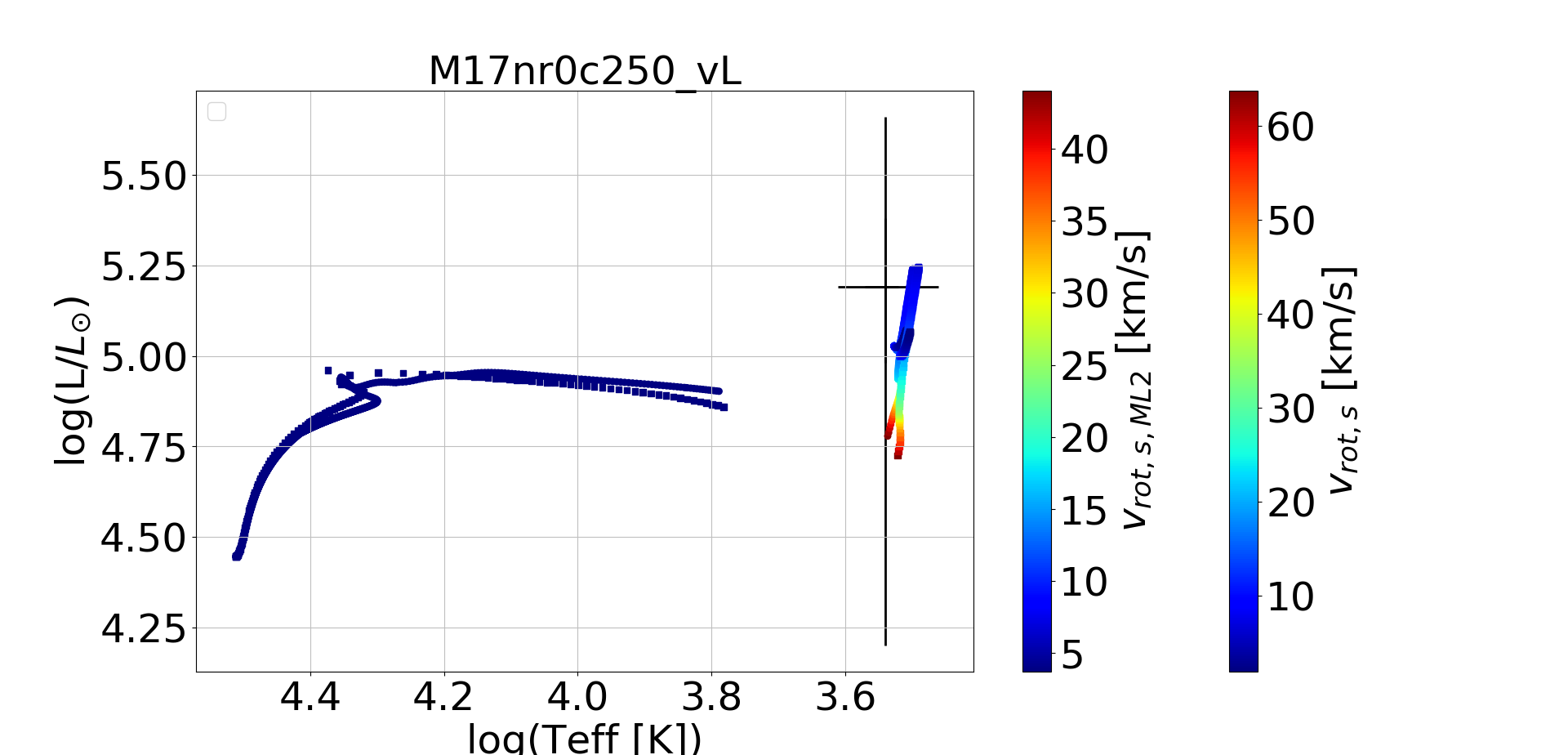}{0.5\textwidth}{}
          }
\gridline{\fig{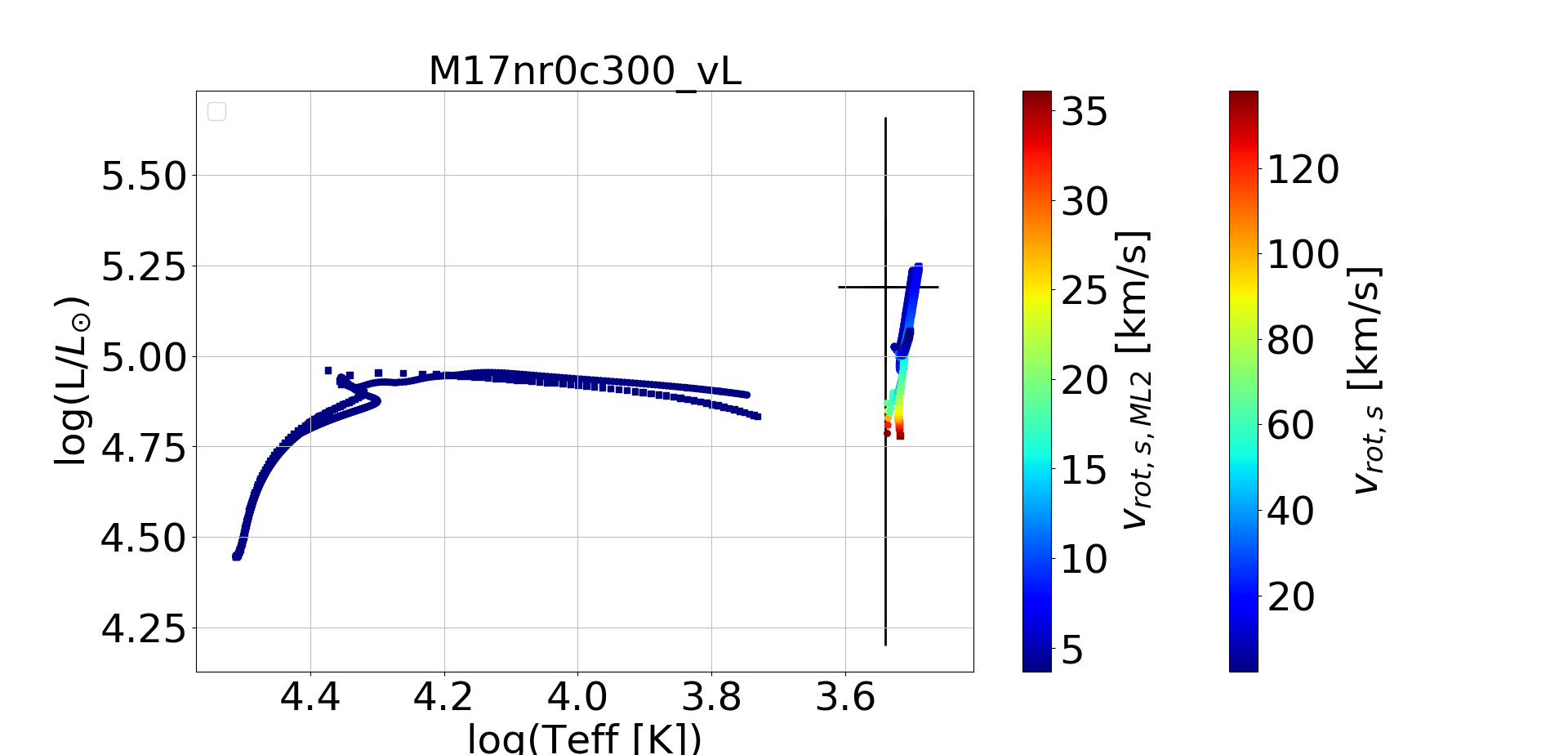}{0.5\textwidth}{}
          \fig{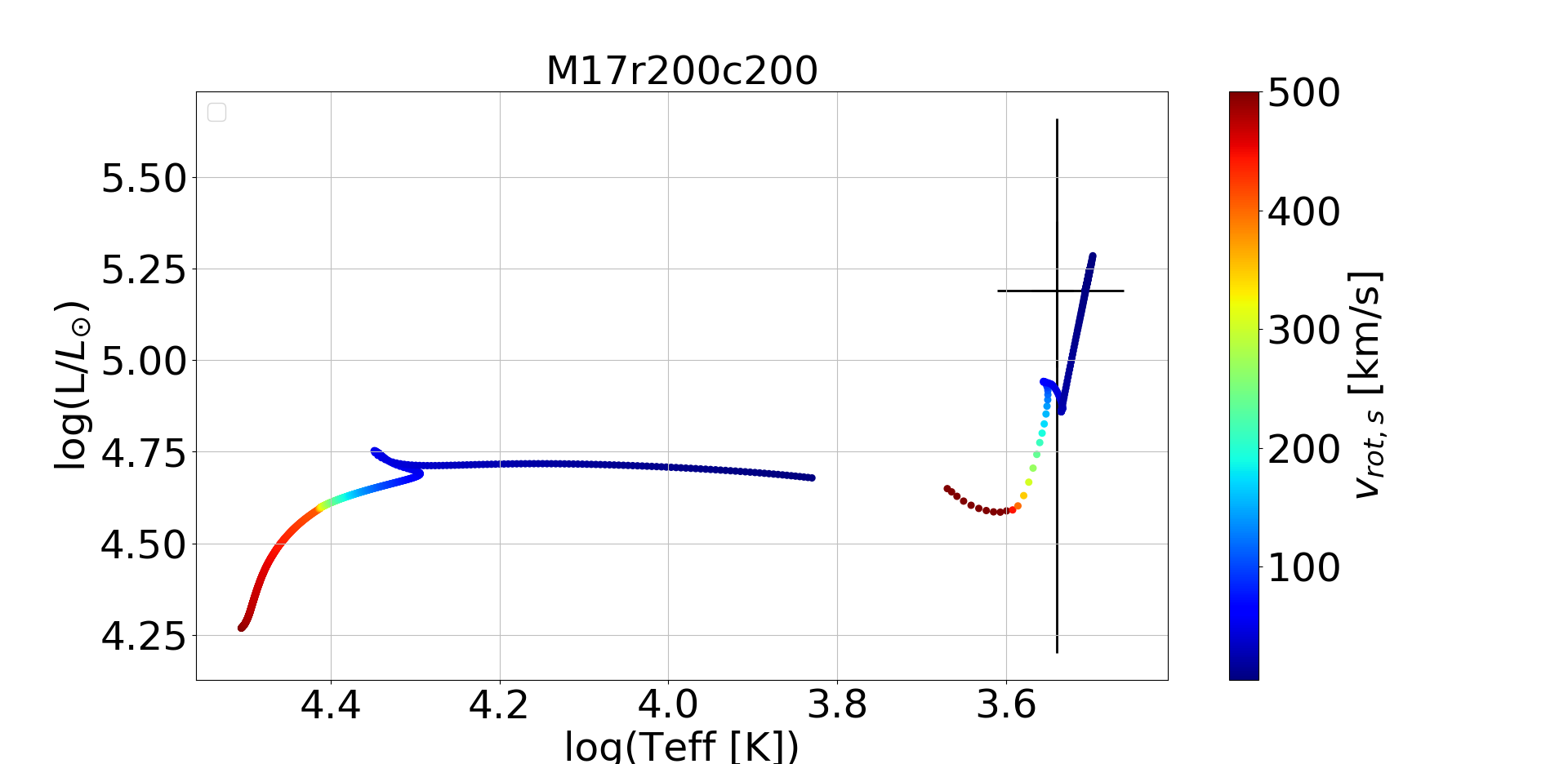}{0.5\textwidth}{}
          }
\gridline{\fig{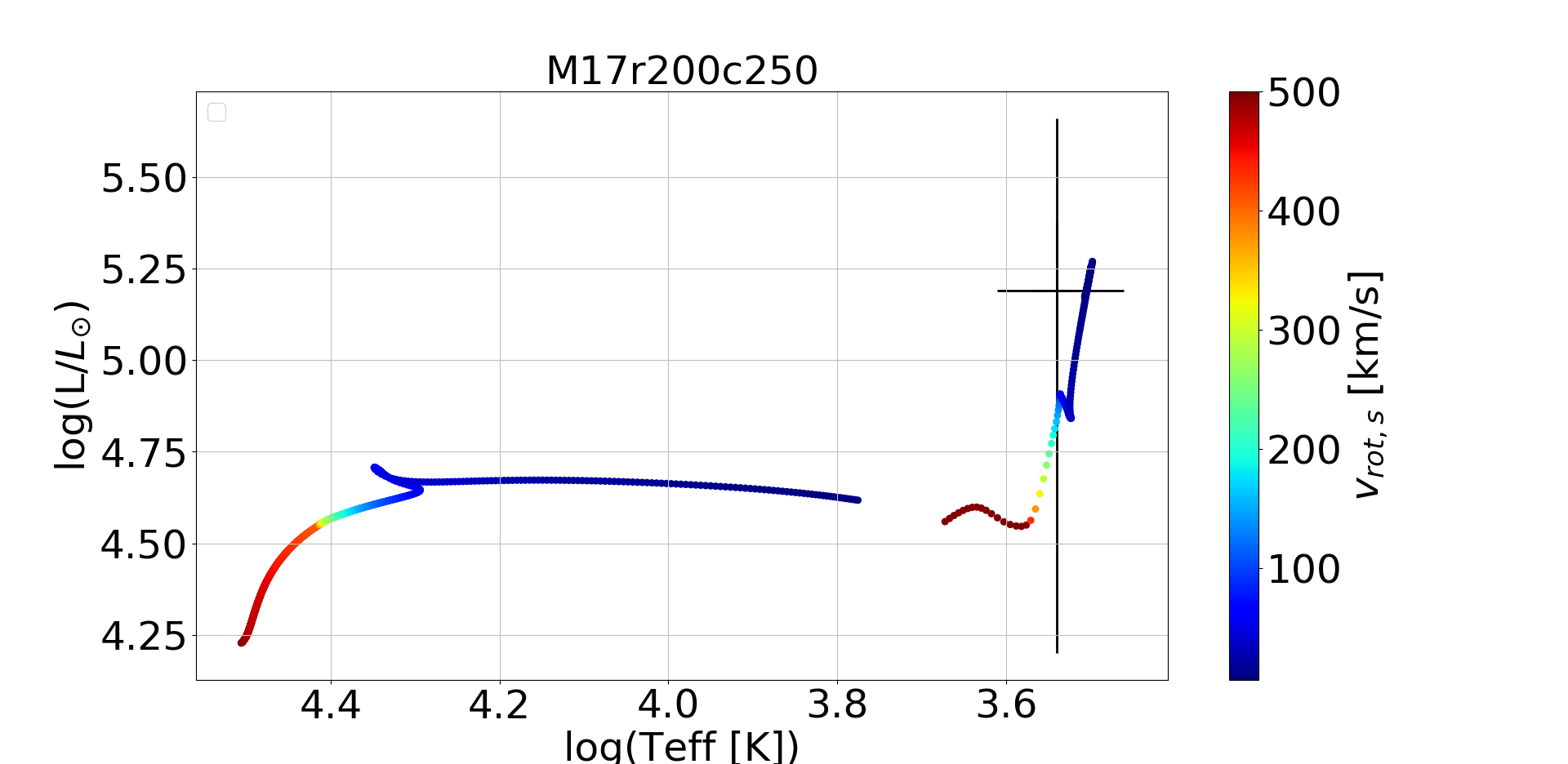}{0.5\textwidth}{}
          \fig{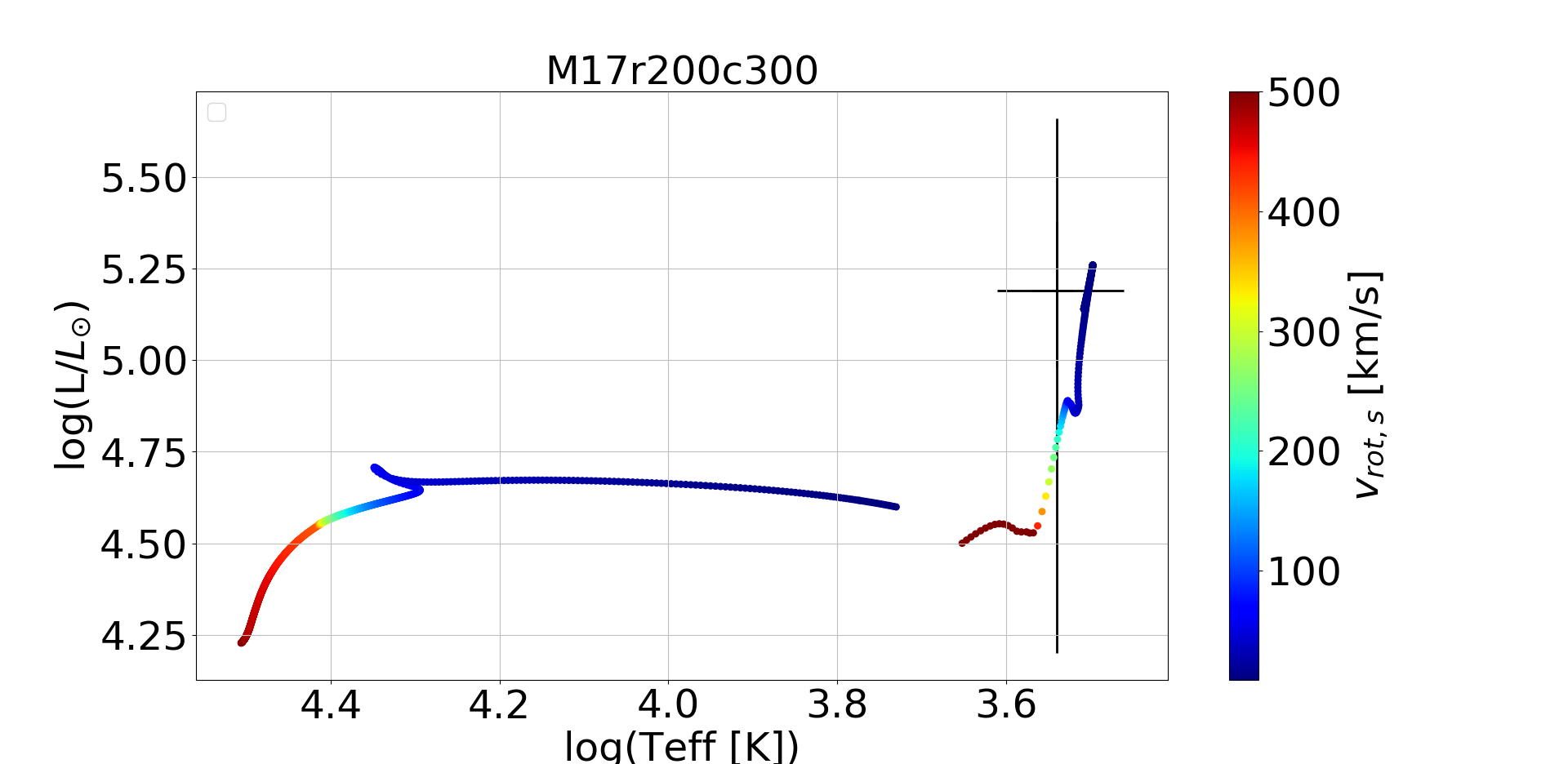}{0.5\textwidth}{}
          }
\caption{Evolution of the non--rotating and rotating 
merger (``M'') models in the Hertzsprung–-Russell diagram for ZAMS primary and secondary mass of 
17~$M_{\odot}$ and 3~$M_{\odot}$ accordingly ($q =$~0.18; {\it left panel}).
The 3-$\sigma$ observed location of Betelgeuse is also marked for comparison.
The colorbar shows the corresponding surface equatorial rotational velocity values
at different evolutionary stages. In the top two and the left middle panels, square symbols denote
the models run with the ``van Loon'' formula for RSG mass--loss.
The middle--left panel also shows the (artificial) HR
evolution during the numerical {\it MESA} ``merger'' procedure. Similar effects during that
stellar engineering phase are seen in \citet{2019MNRAS.482..438M} (i.e. their Figure 4).
\label{fig:M17}}
\end{figure*}

\begin{figure*}
\gridline{\fig{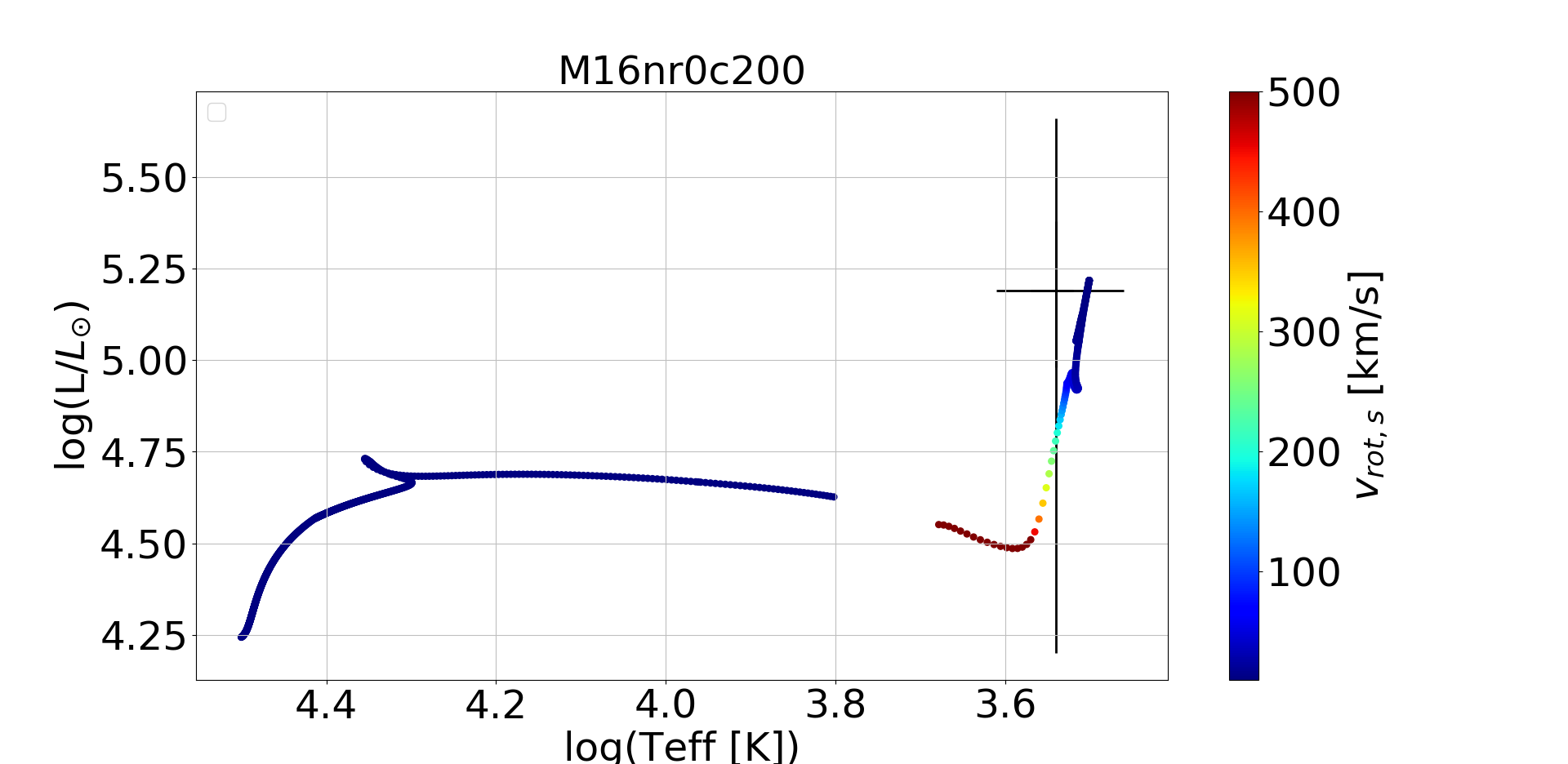}{0.5\textwidth}{}
          \fig{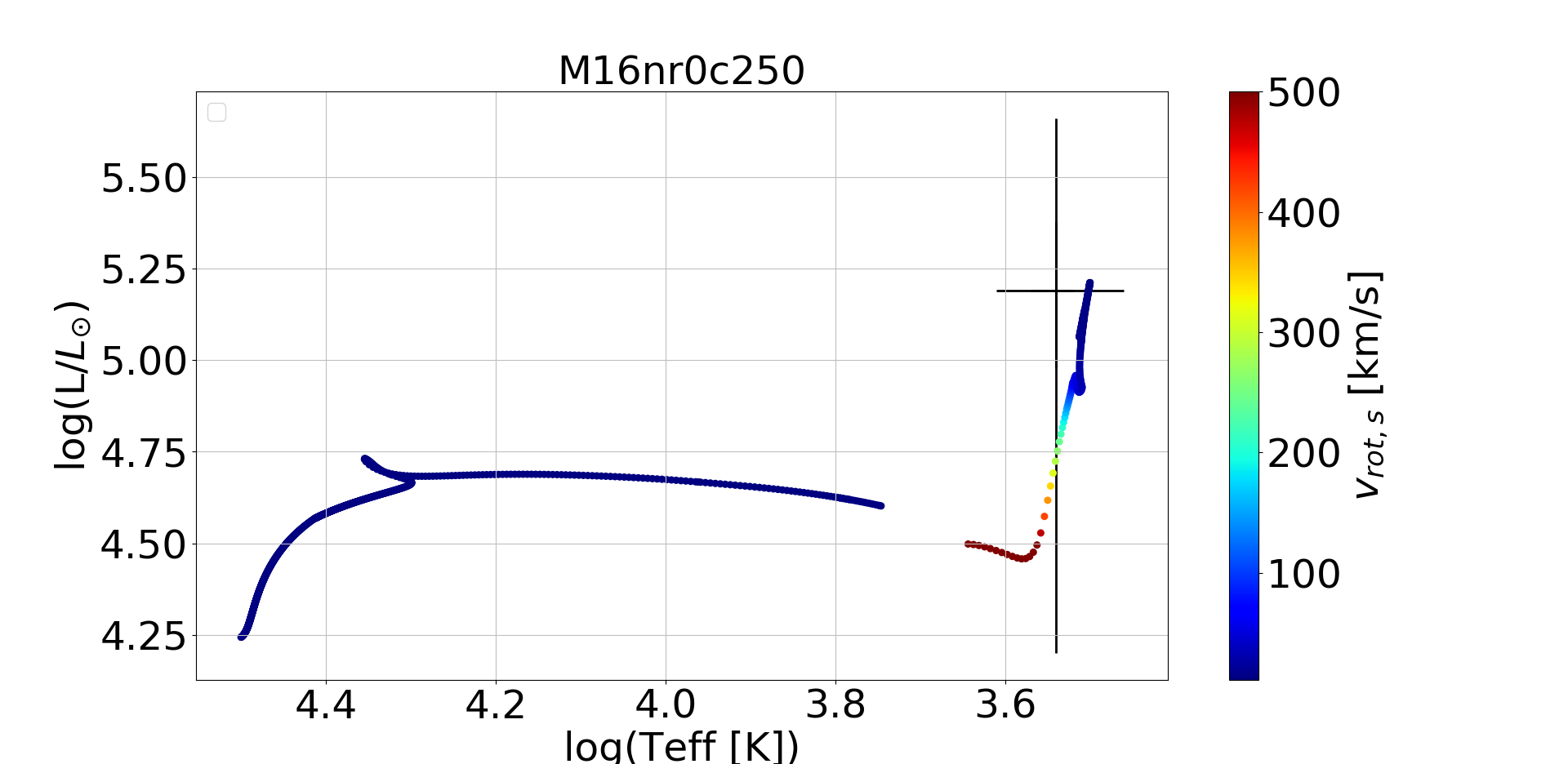}{0.5\textwidth}{}
          }
\gridline{\fig{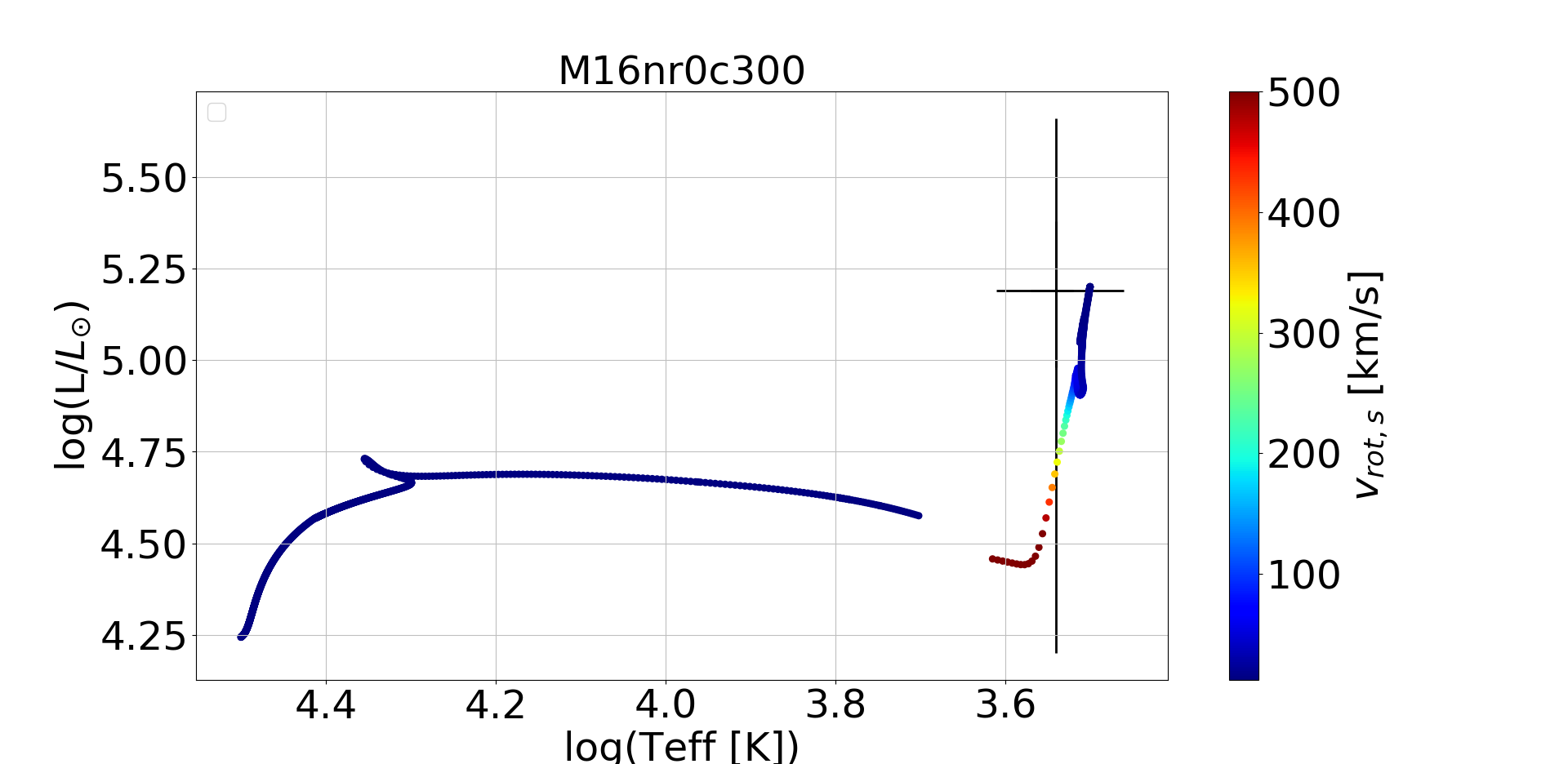}{0.5\textwidth}{}
          \fig{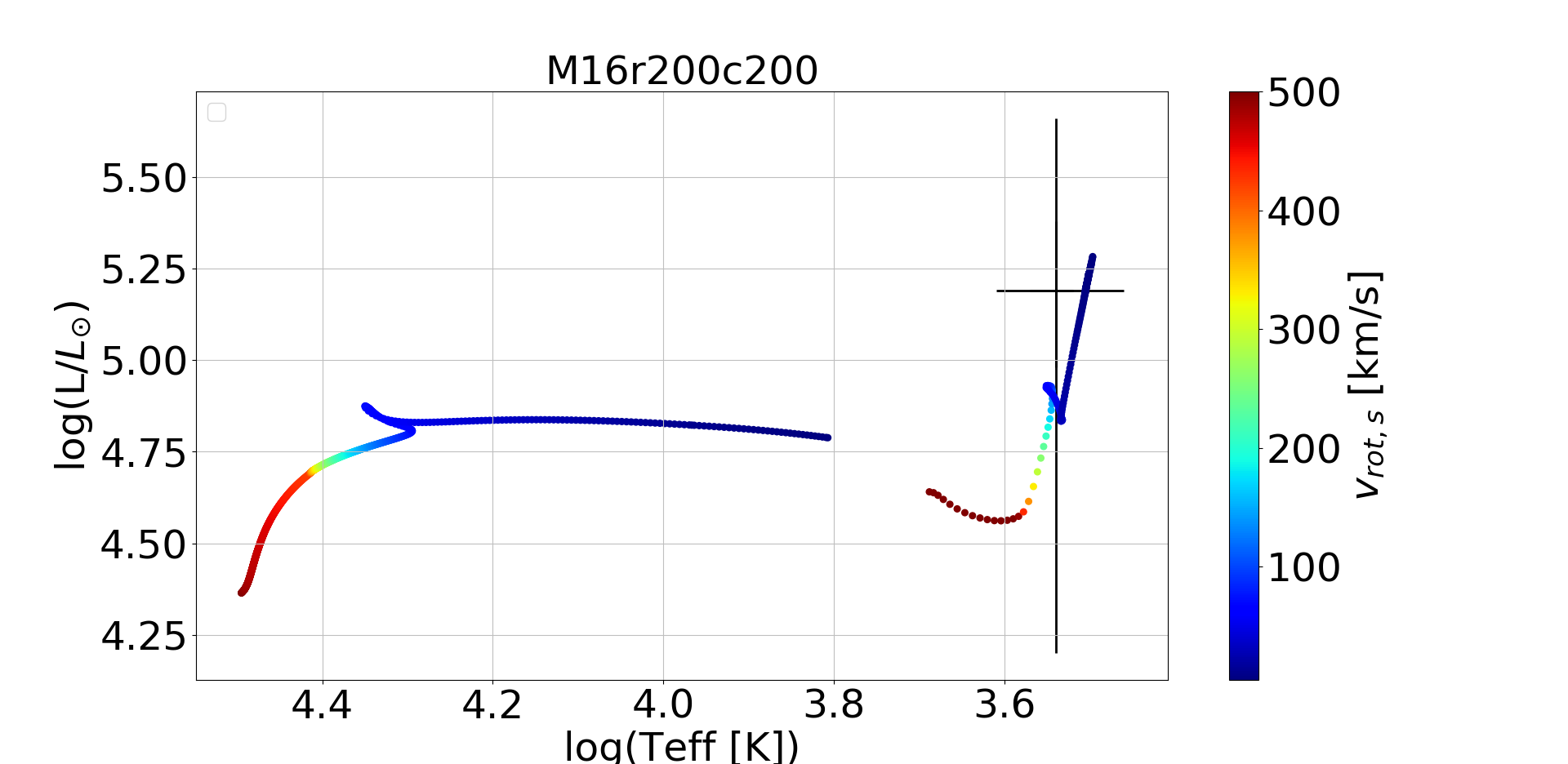}{0.5\textwidth}{}
          }
\gridline{\fig{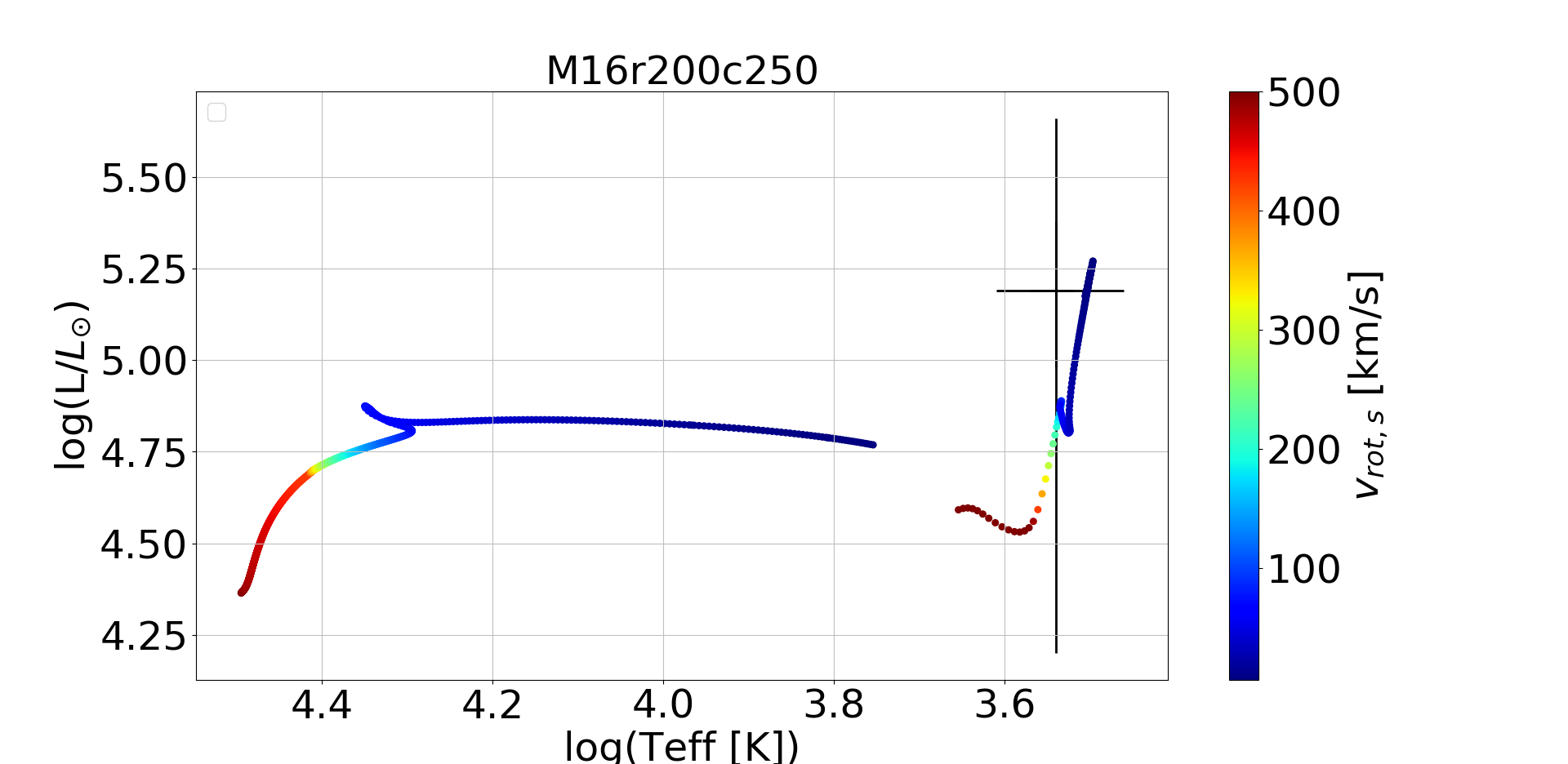}{0.5\textwidth}{}
          \fig{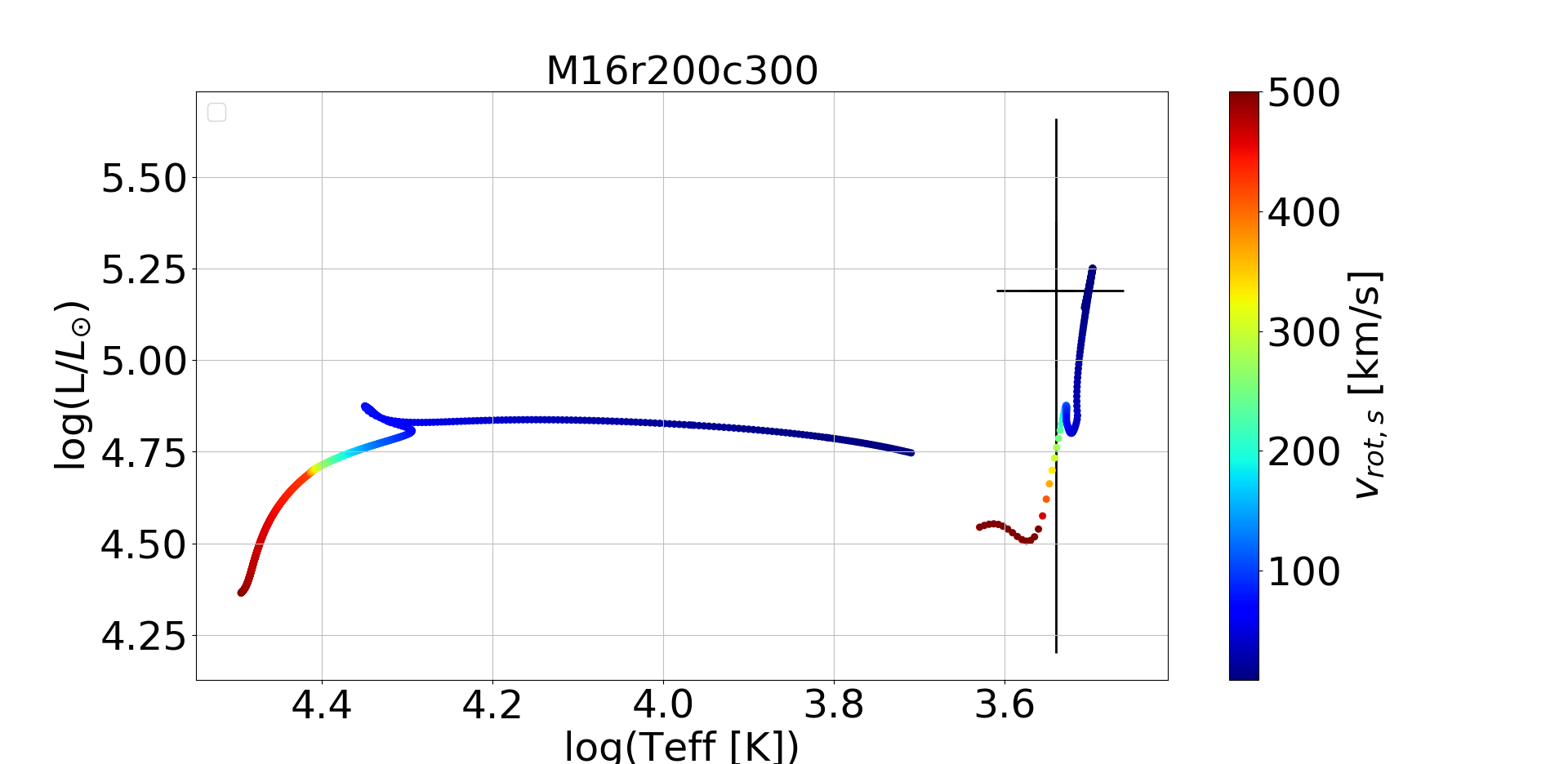}{0.5\textwidth}{}
          }
\caption{Evolution of the non--rotating and rotating 
merger (``M'') models in the Hertzsprung–-Russell diagram for ZAMS primary and secondary mass of 
16~$M_{\odot}$ and 4~$M_{\odot}$ accordingly ($q =$~0.25; {\it left panel}).
The 3-$\sigma$ observed location of Betelgeuse is also marked for comparison.
The colorbar shows the corresponding surface equatorial rotational velocity values
at different evolutionary stages.
\label{fig:M16}}
\end{figure*}

%\begin{figure*}
%\gridline{\fig{20p1.png}{0.5\textwidth}{}
%          \fig{20p2.png}{0.5\textwidth}{}
%          }
%\gridline{\fig{20p3.png}{0.5\textwidth}{}
%          \fig{20p4.png}{0.5\textwidth}{}
%          }
%\caption{Same as in Figure~\ref{fig:M15} but for the 20~$M_{\odot}$ merger (``M'') models.
%\label{fig:M20}}
%\end{figure*}

\begin{figure*}
\begin{center}
\includegraphics[angle=0,width=18cm,trim=0.in 0.25in 0.5in 0.15in,clip]{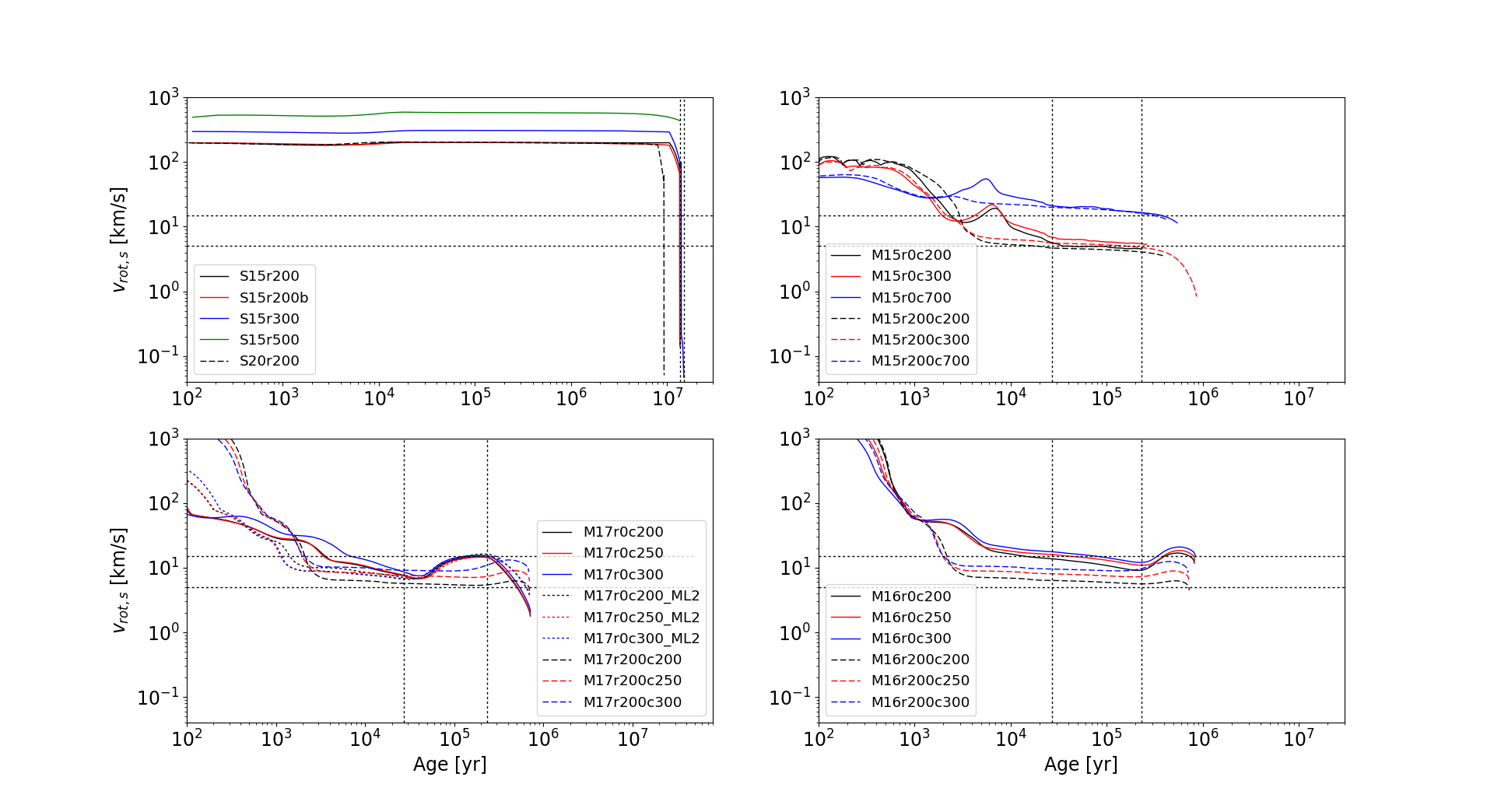}
\caption{Evolution of surface equatorial rotational velocity (v$_{\rm rot,s}$) for 15, 20~$M_{\odot}$ single evolution ({\it upper left panel}) 
and post--merger evolution for primary stars with the following properties:
$\mathcal{M_{\rm 1}} =$~15~$M_{\odot}$, $\mathcal{M_{\rm 2}} =$~1~$M_{\odot}$ ($q =$~0.07; {\it upper right panel}), 
$\mathcal{M_{\rm 1}} =$~17~$M_{\odot}$, $\mathcal{M_{\rm 2}} =$~3~$M_{\odot}$ ($q =$~0.18; {\it lower left panel}) and
$\mathcal{M_{\rm 1}} =$~16~$M_{\odot}$, $\mathcal{M_{\rm 2}} =$~4~$M_{\odot}$ ($q =$~0.25; {\it lower right panel}).
Solid lines correspond to the initally non--rotating primaries and dashed lines to the initially rotating primaries with $v_{\rm 1,rot,i} =$~200~km~$s^{-1}$.
The horizontal dashed lines denote the 5--15~km~s$^{-1}$ range and the vertical
dashed lines correspond to the times when the models are within the 3--$\sigma$ error bars of the observed luminosity and effective temperature
for Betelgeuse. See Table~\ref{T1} for details about the model properties.}
\label{Fig:surf_vel_evol}
\end{center}
\end{figure*}

\subsection{{\it Post--merger structure relaxation with {\it MESA}.}}\label{merger_relax}

To quantitatively evaluate the physical properties and long--term evolution of post--merger objects produced the binary progenitor
systems discussed in Section~\ref{merger_analytical}, we use the open--source stellar evolution code {\it MESA} version 10108. 
More specifically, we take advantage
of the new model relaxation feature described in Appendix B of \citep{2018ApJS..234...34P} that allows the user to import
the entropy (or, alternatively the density/temperature, density/internal energy or pressure/temperature) profile, and the 
specific angular momentum and composition profile of an 1D model into {\it MESA}, regardless of whether it was calculated in an external code
or {\it MESA} itself.
The original model profiles are then gradually numerically ``relaxed''  to the input entropy, angular momentum and composition profiles in {\it MESA}
with the goal to achieve hydrostatic equilibrium computed with an Equation of State (EOS) that is consistent with the one used in the code. Subsequently,
the ``relaxed'' models can be evolved for long, stellar evolution time--scales with all physics modules turned on.

In our analysis, we start by computing the single evolution of massive primaries with the following properties during ZAMS:
$\mathcal{M_{\rm 2}} =$~15, 20~$M_{\odot}$, and initial equatorial rotational velocity $v_{\rm 1,rot,i} =$~0, 200, 300, 500~km~s$^{-1}$ 
therefore accounting for both non--rotating and rapidly rotating pre--merger primary stars. For all models we are using the
Ledoux criterion for convection with the mixing--length theory (MLT) coefficient, $\alpha_{\rm MLT} =$~1.5. Semiconvective and thermohaline
mixing with fiducial coefficients of $\alpha_{\rm semi} =$~0.04 and $\alpha_{\rm th} =$1.0 respectively are included \citep{1980A&A....91..175K}.
We also use the 21--isotope ``approx21.net'' nuclear reaction network \citep{2000ApJS..129..377T} and the ``Helmholtz'' EOS 
\citep{2000ApJS..126..501T}. For the rotating models, angular momentum diffusion and rotationally--induced mixing is 
computed by using the supplied perscriptions \citep{2005ApJ...626..350H} including the effects of magnetic fields due
to Spruit--Taylor dynamo action \citep{2002A&A...381..923S}. 
For each rotational instability (such as meridional (Eddington--Sweet; ``ES'') circulation, dynamical shear instability (``DSI''), seculary shear instability (``SSI''),
Soldberg--Hoiland instability (``SHI''), Goldreich--Schubert--Fricke (``GSF'') instability and Spruit-Taylor (``ST'' dynamo) a viscosity and a diffusion coefficient
are adopted based on standard assumptions for massive stars \citep{2000ApJ...528..368H,2002A&A...381..923S,2003ApJ...591..288H}.
{\it MESA} allows the user to set the efficiency of each mechanism (which we take to be 1.0) separately, and then computes each contribution to the total viscosity
and diffusion coefficient independently as described in \citet{2000ApJ...528..368H,2005ApJ...626..350H}.
These coefficients are then used to calculate the total viscosity and diffusion coefficients that determine the efficiency 
of angular momentum mixing in different regions of the stars by solving the angular momentum diffusion equation in 1D spherical coordinates.
We note that these angular momentum mixing processes were adopted for all of our rotating models (both the single rotating models and the rotating
post--merger models). As a consequence, our results on the evolution of rotation in these models may be dependent on how efficient angular momentum
mixing is in massive stars, which is an actively debated topic involving several poorly constrained parameters. Alternative angular momentum mixing
processes (see, for example \citealt{2014ApJ...796...17F}) may therefore impact our results.
Steady--state radiatively--driven mass--loss is also computed following \citet{2001A&A...369..574V}. Temperature--dependent 
mass--loss during the RSG phase is based on the ``de Jager'' prescriptions \citep{1988BICDS..35..141D}. We also run a subset of models
for the $q =$~0.18 case ($\mathcal{M_{\rm 1}} =$~17~$M_{\odot}$) using the ``van Loon'' formula \citep{2005A&A...438..273V} suggesting
stronger winds during the RSG phase \citep{2014ARA&A..52..487S} in order to investigate the effect of mass--loss on the final
surface rotation rate.

We computed two sets of models: the ``Single (S)'' and the ``Merged (M)'' group. The ``S'' models refer to the evolution of massive,
rotating or non--rotating single stars that do not suffer a merger. We run a set of ``S'' models in order to calculate the evolution
of the surface equatorial rotational velocity of massive stars and determine their rotation rates after the end of their main--sequence
lifetimes, during the supergiant phases of their evolution. This is equivalent to the work done by \citep{2017MNRAS.465.2654W}
with the purpose to determine for how long single massive supergiant stars rotate with speeds consistent with those reported for Betelgeuse.
The ``M'' group of models is essentially a sub--sample of the ``S'' group but with the key difference that we stop the single evolution
part of the computation during the ``Hertzsprung--gap'' crossing phase while the primary is expanding toward the supergiant branch after
the end of the main--sequence. It is during that phase that we assume the first contact with secondary occurs triggering
the merger process. In our analysis we consider different values for the initial contact radius for primary radii
$R_{\rm 1} =$~200, 250, 300 or 700~$R_{\odot}$. 
From hereafter we adopt the notations $<$X$>$$<$Y$>$r$<$Z$>$ and $<$X$>$$<$Y$>$r$<$Z$>$c$<$V$>$ where X$=${S,M} for the ``Single'' or
the ``Merged'' case, Y$= \mathcal{M_{\rm 1}}$ in solar masses, Z$=v_{\rm rot, 0}$ (ZAMS surface equatorial rotation rate) in 
km~s$^{-1}$ and V$=R_{\rm 1}$ in $R_{\odot}$. Details for each models are given extensively in Table~\ref{T1}. As an example, model ``M17r200c300'' 
corresponds to a $\mathcal{M_{\rm 1}} =$~17~$M_{\odot}$ primary rotating at 200~km~s$^{-1}$ during ZAMS that suffered a merger with an 4~$M_{\odot}$ 
secondary after the end of the main--sequence and at a radius of $R_{\rm 1} =$~300~$R_{\odot}$.  We also run a model where we ignore the effects
of angular momentum transfer due to magnetic fields (model ``S15r200b'') and a model with enhanced mass--loss rate during the main--sequence
(model ``S20r0b'', where we set the mass--loss efficiency coefficients ($\eta_{\rm ML}$) to 5 times higher than the default value of 1.0) 
to investigate the corresponding effects on the evolution of surface rotation rate.

To compute the post--merger evolution of the ``M'' set of models, we adjust their specific angular momentum profiles for different choices of contact radii $R_{\rm 1}$, 
according to Equations~\ref{Eq:js2} and~\ref{Eq:js3} by using the {\it MESA} model relaxation feature. 
In addition, following arguments discussed in Section~\ref{ICs} we assume that the post--merger composition and entropy profiles remain unaffected.
Figure~\ref{Fig:15sm_mapping} shows the pre-- and post--merger internal distribution of specific angular momentum,
angular velocity, density and temperature for the ``M15r200c200'' models. It can be seen that the envelope of the primary 
experiences a significant spin--up due to the spiral--in of the 1~$M_{\odot}$ companion star. Depending on the specifics of re--distribution of angular momentum due
to diffusive processes and convection in the envelope of the primary, this high rotation rate may persist for long time--scales and up to the supergiant stage
of the post--merger evolution. We investigate this in the following paragraph.

\subsection{{\it Post--merger evolution with {\it MESA}.}}\label{merger_evol}

The long--term evolution of both the ``S'' and ``M'' class of models is followed for up to the supergiant, core He--burning, phases
providing us with ``Hertzsprung--Russell'' (HR) tracks and predictions about the evolution of surface equatorial rotational velocity that
can be directly compared against observations of rapidly--rotating supergiant stars and, more specifically, Betelgeuse. 

Betelgeuse has been discussed as a candidate of a supergiant star that has recently suffered a merger event with a smaller companion 
\citep{2017MNRAS.465.2654W,2018MNRAS.479..251N}. We adopt the observational results of \citet{2016ApJ...819....7D} indicating
$L/L_{\odot} =1.3^{+0.7}_{-0.5} \times 10^{5}$, $R/R_{\odot} = 887 \pm 203$ and $T_{\rm eff} = 3500 \pm 200$~K where the uncertainties 
in luminosity and radius are dominated by the uncertainty to the distance of Betelgeuse, $d = 197 \pm 45$~pc \citep{2008AJ....135.1430H,2017AJ....154...11H}.
In addition, \citet{1996ApJ...463L..29G,1998AJ....116.2501U,2018A&A...609A..67K} report surface rotational velocities in the range $\sim$~5--15~km~s$^{-1}$.

\section{Results}\label{results}

Table~\ref{T1} details the properties of the models studied in this work. $\log(L_{\rm 0}/L_{\odot})$, $\log T_{\rm eff, 0}$ and $v_{\rm rot, 0} $ 
correspond to the values of the luminosity (in solar units), effective temperature and surface equatorial rotational velocity 
of each model during the supergiant phase respectively. $\Delta t_{\rm target}$ denotes the time (in years) that each model spends within
the range of target observed values for Betelgeuse (5--15~km~s$^{-1}$). 
Figure~\ref{fig:single} shows the HR tracks of the ``S'' models and Figures~\ref{fig:M15},~\ref{fig:M17} and ~\ref{fig:M16}
the HR tracks of the $\mathcal{M_{\rm 1}} =$~15, 17 and 16~$M_{\odot}$ ``M'' models respectively, including a colorscale that denotes the surface equatorial
velocity at different stages in units of km~s$^{-1}$. The small gaps in data between the pre--merger and the post--merger evolution in the HR plots
correspond to the first few steps of numerical relaxation that we omitted from the plot. The post--merger evolution is therefore plotted starting at
the first dynamically stable, relaxed state of the post--merger object.
Figure~\ref{Fig:surf_vel_evol} shows the evolution of surface equatorial rotational velocity
for rotating ``S'' models and all ``M'' models after merger. 

All the ``S'' models evolve to a red supergiant phase within $\sim$~10--13 million years with the exception of the most rapidly--rotating model ``S15r500'' that evolves to
a blue supergiant that is inconsistent with the observations of Betelgeuse, and as such we exclude it from further discussion. 
The luminosity and effective temperatures of the ``S'' models that evolve to the red supergiant phase are well within the error bars of those of Betelgeuse. 
However, after the end of the main--sequence and during their thermal expansion
toward the supergiant phase, angular momentum conservation leads to a significant spin--down and surface rotation velocities $\sim$~0.2~km~s$^{-1}$ for the 
rotating 15~$M_{\odot}$ ``S'' models and  $\sim$~0.06~km~s$^{-1}$ for the rotating 20~$M_{\odot}$ model respectively. The exclusion of the effects
of angular momentum transport due to magnetic fields (model ``S15r200b'') and enhanced main--sequence steady--state mass--loss rates (model ``S20r0b'')
do not significantly affect the predictions for $\log(L_{\rm 0}/L_{\odot})$, $\log T_{\rm eff, 0}$ and $v_{\rm rot, 0} $. During the ascent toward higher luminosities and
the supergiant phase the surface equatorial rotational velocity gradually declines for all ``S'' models due to the envelope expansion but only spends a very short
amount of time ($\sim$ a few to tens of years in some cases) within the target 5--15~km~s$^{-1}$ range. That is consistent with the findings
of \citealt{2017MNRAS.465.2654W}; (See their Figure 7 for more information). This result indicates that models of single stellar evolution with rapid rotation
at the ZAMS do not accurately predict the observed characteristics of Betelgeuse. 

All of the models in the ``M'' subset evolve toward the observed, 3--$\sigma$ luminosity and effective temperature for Betelgeuse during their post--merger
evolution. In Figures ~\ref{fig:M15},~\ref{fig:M17} and~\ref{fig:M16} the colorscale corresponds to the magnitude of surface equatorial rotational velocity 
in units of km~s$^{-1}$. A
significant spin--up up to $\sim$~100~km~s$^{-1}$ is acquired within a thermal adjustement timescale following the merger. {\it MESA} is numerically 
``relaxing'' to an acceptable model for a time period less than the thermal adjustement timescale (see HR track for model ``M17nr0c300'').
Fast rotation is also maintained all throughout the ascent to the supergiant phase, with $v_{\rm rot, 0} >$~5~km~s$^{-1}$ during the time of 
peak luminosity as indicated in their corresponding HR tracks. 
As can be seen in Figure~\ref{Fig:surf_vel_evol}, these rapid rotation rates persist for hundreds of thousands of years for the ``M'' models, 
making it possible to detect rapidly--rotating supergiant stars in higher rates. 
Higher RSG mass loss rates using the ``van Loon'' formula  (the ``M17r0c$<$V$>$\_vL'' models) do not appear to significantly reduce the final surface
rotational velocities.
The preservation of high rotation for long time--scales is 
predominantly due to a fully--convective stellar envelope for $r > 1.54 \times 10^{12}$~cm that assists in establishing near solid--body rotation. It is worth cautioning,
however, that this is the outcome assuming standard simplified 1D parametrizations of angular momentum transport, a process that is inherently three--dimensional,
and not fully understood especially for massive stars as it depends on both internal structure conditions as well as binary interaction effects.
Finally, the results detailed in Table~\ref{T1} for the ``M'' suite of models show that for primaries of the same mass, and for the same contact radius, the
inclusion of pre--existing rotation for the primary does not alter the results significantly. 

\section{Discussion}\label{disc}

In this paper we study the evolution -- including the angular momentum evolution -- of low mass--ratio mergers (0.07$<q<$0.25) where a massive primary
star engulfs its smaller ($\mathcal{M_{\rm 2}} \sim$~1--4~$M_{\odot}$) companion during their thermal expansion phase after the end of the main sequence. 
We devise the initial conditions for the binary progenitor system and an approximate, analytical model that characterizes the deposition of
angular momentum in the envelope of a massive primary during the spiraling--in of the secondary and up to the point when the secondary
becomes tidally disrupted around the He core of the primary. We then calculate the perturbed, post--merger specific angular momentum
distribution and use it to simulate the post--merger evolution with the stellar evolution code {\it MESA}. We consider two sets of models,
one where we calculated the evolution of single, rapidly--rotating stars (the ``S'' group) and another where massive stars suffer a merger event in order
to predict the evolution of surface equatorial rotational velocity toward the supergiant phase (the ``M'' group) 
and compare our results with the observations of the star Betelgeuse.

Our analysis indicates that low mass--ratio ($q =\mathcal{M}_{\rm 2}/\mathcal{M}_{\rm 1} =$~0.07--0.25) mergers initiated during the post--main sequence 
evolution of a primary with mass 15--17~$M_{\odot}$ and when the primary
reaches a radius of $\sim$200--300~$R_{\odot}$ evolve to red supergiant stars that are rapid rotators for up to $\sim$200,000~years, in good agreement with the
observations of Betelgeuse. In Section~\ref{merger_necessary} we considered also the possibility that Betelgeuse might have been spun--up by accretion and then
released in a supernova explosion of its companion. A survey of existing population synthesis and binary stellar evolution calculations shows that
there is no clear and compelling path to the desired result, because the outcomes are extremely sensitive to initial conditions and model assumptions.
On the other hand, we are unable at this point to rule out the accretion option for Betelgeuse. This may require further investigation in a future paper.
In any case, a two--step process may be required to obtain both the observed runaway velocity and the high equatorial rotational velocity
of Betelgeuse: a) the dynamical ejection of a binary from Betelgeuse's birth cluster as the origin of its observed space velocity (with binary parameters that allow the
system to survive the ejection) and b) the subsequent merger or accretion--induced spin--up followed by a supernova explosion.

We have also established via binary system energetics arguments and references to relevant past N--body simulation studies, that 
young dense stellar clusters can successfully eject intact runaway binary systems (with ejection velocities $>$~30~km~s$^{-1}$) with 
properties that allow the primary to obtain a non--canonical, rapid surface equatorial velocity following an early Case B merger that occurs
after its terminal age main sequence and during its Hertzsprung gap--crossing phase. Furthermore, we have estimated that such ``quiet'' mergers will 
not lead to significant rejuvenation of the primary and the formation of a BSG star akin to the progenitor of SN 1987A, but will likely yield rapidly 
rotating RSGs instead since the secondary will not penetrate deep into the core of the primary but be deposited on the H--burning shell.

It should be emphasized that some of our results may depend on the details of angular momentum transport in the post merger object and the mechanisms
of angular momentum mixing that are available in the {\it MESA} code via 1D parametrized perscriptions, where we adopt standard values pertaining
to massive stars \citep{2003ApJ...591..288H,2005ApJ...626..350H,2012A&A...537A.146E}. Our results can be explored for other mechanisms of internal
angular momentum not included in this work (see, for example \citealt{2014ApJ...796...17F} but that is beyond the scope of this paper.

These results indicate that stellar mergers, a natural biproduct of stellar duplicity, is a prevalent feature in the Universe \citep{2012Sci...337..444S,2014ApJ...782....7D}, 
is a reasonable formation channel for rapidly--rotating giant and supergiant stars \citep{2015ApJ...807...82T} and a mechanism that affects both the
 details of stellar evolution but also the properties of circumstellar environments. 
Even though our approach was approximate and utilized many simplifying assumptions, this work serves as a proof of principle that
certain merger scenarios can reproduce observations of a subset of giant and supergiant stars, and potentially other categories of stars
that are generally considered ``outliers''.

In the near future we aim to perform realistic, 3D simulations of dynamical mergers with the same initial properties as the ones discussed in this work with the aim
to more accurately assess the efficiency of angular momentum deposition but also better quantify the mechanically--driven mass--loss that may lead to the
formation of circumstellar mass outflows. We also aim to more carefully investigate the effect of these merger processes 
on nucleosynthesis, and the potential to detect relevant signatures in the spectra of candidate stars.
These calculations will allow us to obtain a better model for the structure of the post--merger objects
that can then be spherically--averaged and evolved with {\it MESA} with the aim to 
further constrain the long--term effects of binary coalescence in massive stellar evolution.

\acknowledgments

We would like to thank J. Craig Wheeler, Pavel Kroupa and Pablo Marchant 
for useful discussions and comments. We would also like to thank our
referee for the very constructive criticism provided that helped us
improve the quality of our paper.
We wish to acknowledge the support from the National Science 
Foundation through CREATIV grant AST–-1240655.
EC would like to thank the National Science Foundation for its support
through award number AST--1907617 and the Louisiana State University College of
Science and the Department of Physics \& Astronomy for their support.

\software
{\it MESA} \citep{2015ApJS..220...15P}
{\tt Matplotlib} \citep{Hunter2007}.

\bibliography{refs}

\begin{thebibliography}{}
\expandafter\ifx\csname natexlab\endcsname\relax\def\natexlab#1{#1}\fi
\providecommand{\url}[1]{\href{#1}{#1}}
\providecommand{\dodoi}[1]{doi:~\href{http://doi.org/#1}{\nolinkurl{#1}}}
\providecommand{\doeprint}[1]{\href{http://ascl.net/#1}{\nolinkurl{http://ascl.net/#1}}}
\providecommand{\doarXiv}[1]{\href{https://arxiv.org/abs/#1}{\nolinkurl{https://arxiv.org/abs/#1}}}

\bibitem[{{Bally}(2008)}]{2008hsf1.book..459B}
{Bally}, J. 2008, {Overview of the Orion Complex}, ed. B.~{Reipurth}, Vol.~4,
  459

\bibitem[{{Blaauw}(1961)}]{1961BAN....15..265B}
{Blaauw}, A. 1961, \bain, 15, 265

\bibitem[{{Blaauw}(1964)}]{1964ARA&A...2..213B}
---. 1964, \araa, 2, 213, \dodoi{10.1146/annurev.aa.02.090164.001241}

\bibitem[{{Brice{\~n}o} {et~al.}(2005){Brice{\~n}o}, {Calvet}, {Hern{\'a}ndez},
  {Vivas}, {Hartmann}, {Downes}, \& {Berlind}}]{2005AJ....129..907B}
{Brice{\~n}o}, C., {Calvet}, N., {Hern{\'a}ndez}, J., {et~al.} 2005, \aj, 129,
  907, \dodoi{10.1086/426911}

\bibitem[{{Brott} {et~al.}(2011{\natexlab{a}}){Brott}, {de Mink}, {Cantiello},
  {Langer}, {de Koter}, {Evans}, {Hunter}, {Trundle}, \&
  {Vink}}]{2011A&A...530A.115B}
{Brott}, I., {de Mink}, S.~E., {Cantiello}, M., {et~al.} 2011{\natexlab{a}},
  \aap, 530, A115, \dodoi{10.1051/0004-6361/201016113}

\bibitem[{{Brott} {et~al.}(2011{\natexlab{b}}){Brott}, {Evans}, {Hunter}, {de
  Koter}, {Langer}, {Dufton}, {Cantiello}, {Trundle}, {Lennon}, {de Mink},
  {Yoon}, \& {Anders}}]{2011A&A...530A.116B}
{Brott}, I., {Evans}, C.~J., {Hunter}, I., {et~al.} 2011{\natexlab{b}}, \aap,
  530, A116, \dodoi{10.1051/0004-6361/201016114}

\bibitem[{{Brown} {et~al.}(1994){Brown}, {de Geus}, \& {de
  Zeeuw}}]{1994A&A...289..101B}
{Brown}, A.~G.~A., {de Geus}, E.~J., \& {de Zeeuw}, P.~T. 1994, \aap, 289, 101.
\newblock \doarXiv{astro-ph/9403051}

\bibitem[{{Ceillier} {et~al.}(2017){Ceillier}, {Tayar}, {Mathur}, {Salabert},
  {Garc{\'\i}a}, {Stello}, {Pinsonneault}, {van Saders}, {Beck}, \&
  {Bloemen}}]{2017A&A...605A.111C}
{Ceillier}, T., {Tayar}, J., {Mathur}, S., {et~al.} 2017, \aap, 605, A111,
  \dodoi{10.1051/0004-6361/201629884}

\bibitem[{{Claret} \& {Gimenez}(1989)}]{1989A&AS...81...37C}
{Claret}, A., \& {Gimenez}, A. 1989, \aaps, 81, 37

\bibitem[{{Costa} {et~al.}(2015){Costa}, {Canto Martins}, {Bravo},
  {Paz-Chinch{\'o}n}, {das Chagas}, {Le{\~a}o}, {Pereira de Oliveira},
  {Rodrigues da Silva}, {Roque}, {de Oliveira}, {Freire da Silva}, \& {De
  Medeiros}}]{2015ApJ...807L..21C}
{Costa}, A.~D., {Canto Martins}, B.~L., {Bravo}, J.~P., {et~al.} 2015, \apjl,
  807, L21, \dodoi{10.1088/2041-8205/807/2/L21}

\bibitem[{{de Jager} {et~al.}(1988){de Jager}, {Nieuwenhuijden}, \& {van der
  Hucht}}]{1988BICDS..35..141D}
{de Jager}, C., {Nieuwenhuijden}, H., \& {van der Hucht}, K.~A. 1988, Bulletin
  d'Information du Centre de Donnees Stellaires, 35, 141

\bibitem[{{de Mink} {et~al.}(2013){de Mink}, {Langer}, {Izzard}, {Sana}, \& {de
  Koter}}]{2013ApJ...764..166D}
{de Mink}, S.~E., {Langer}, N., {Izzard}, R.~G., {Sana}, H., \& {de Koter}, A.
  2013, \apj, 764, 166, \dodoi{10.1088/0004-637X/764/2/166}

\bibitem[{{de Mink} {et~al.}(2014){de Mink}, {Sana}, {Langer}, {Izzard}, \&
  {Schneider}}]{2014ApJ...782....7D}
{de Mink}, S.~E., {Sana}, H., {Langer}, N., {Izzard}, R.~G., \& {Schneider},
  F.~R.~N. 2014, \apj, 782, 7, \dodoi{10.1088/0004-637X/782/1/7}

\bibitem[{{Dolan} {et~al.}(2016){Dolan}, {Mathews}, {Lam}, {Quynh Lan},
  {Herczeg}, \& {Dearborn}}]{2016ApJ...819....7D}
{Dolan}, M.~M., {Mathews}, G.~J., {Lam}, D.~D., {et~al.} 2016, \apj, 819, 7,
  \dodoi{10.3847/0004-637X/819/1/7}

\bibitem[{{Dufton} {et~al.}(2011){Dufton}, {Dunstall}, {Evans}, {Brott},
  {Cantiello}, {de Koter}, {de Mink}, {Fraser}, {H{\'e}nault-Brunet},
  {Howarth}, {Langer}, {Lennon}, {Markova}, {Sana}, \&
  {Taylor}}]{2011ApJ...743L..22D}
{Dufton}, P.~L., {Dunstall}, P.~R., {Evans}, C.~J., {et~al.} 2011, \apjl, 743,
  L22, \dodoi{10.1088/2041-8205/743/1/L22}

\bibitem[{{Dufton} {et~al.}(2013){Dufton}, {Langer}, {Dunstall}, {Evans},
  {Brott}, {de Mink}, {Howarth}, {Kennedy}, {McEvoy}, {Potter},
  {Ram{\'{\i}}rez-Agudelo}, {Sana}, {Sim{\'o}n-D{\'{\i}}az}, {Taylor}, \&
  {Vink}}]{2013A&A...550A.109D}
{Dufton}, P.~L., {Langer}, N., {Dunstall}, P.~R., {et~al.} 2013, \aap, 550,
  A109, \dodoi{10.1051/0004-6361/201220273}

\bibitem[{{Dunstall} {et~al.}(2015){Dunstall}, {Dufton}, {Sana}, {Evans},
  {Howarth}, {Sim{\'o}n-D{\'{\i}}az}, {de Mink}, {Langer}, {Ma{\'{\i}}z
  Apell{\'a}niz}, \& {Taylor}}]{2015A&A...580A..93D}
{Dunstall}, P.~R., {Dufton}, P.~L., {Sana}, H., {et~al.} 2015, \aap, 580, A93,
  \dodoi{10.1051/0004-6361/201526192}

\bibitem[{{Eggleton}(1983)}]{1983ApJ...268..368E}
{Eggleton}, P.~P. 1983, \apj, 268, 368, \dodoi{10.1086/160960}

\bibitem[{{Ekstr{\"o}m} {et~al.}(2008){Ekstr{\"o}m}, {Meynet}, {Chiappini},
  {Hirschi}, \& {Maeder}}]{2008A&A...489..685E}
{Ekstr{\"o}m}, S., {Meynet}, G., {Chiappini}, C., {Hirschi}, R., \& {Maeder},
  A. 2008, \aap, 489, 685, \dodoi{10.1051/0004-6361:200809633}

\bibitem[{{Ekstr{\"o}m} {et~al.}(2012){Ekstr{\"o}m}, {Georgy}, {Eggenberger},
  {Meynet}, {Mowlavi}, {Wyttenbach}, {Granada}, {Decressin}, {Hirschi},
  {Frischknecht}, {Charbonnel}, \& {Maeder}}]{2012A&A...537A.146E}
{Ekstr{\"o}m}, S., {Georgy}, C., {Eggenberger}, P., {et~al.} 2012, \aap, 537,
  A146, \dodoi{10.1051/0004-6361/201117751}

\bibitem[{{Ferrario} {et~al.}(2009){Ferrario}, {Pringle}, {Tout}, \&
  {Wickramasinghe}}]{2009MNRAS.400L..71F}
{Ferrario}, L., {Pringle}, J.~E., {Tout}, C.~A., \& {Wickramasinghe}, D.~T.
  2009, \mnras, 400, L71, \dodoi{10.1111/j.1745-3933.2009.00765.x}

\bibitem[{{Fuller} {et~al.}(2014){Fuller}, {Lecoanet}, {Cantiello}, \&
  {Brown}}]{2014ApJ...796...17F}
{Fuller}, J., {Lecoanet}, D., {Cantiello}, M., \& {Brown}, B. 2014, \apj, 796,
  17, \dodoi{10.1088/0004-637X/796/1/17}

\bibitem[{{Gies} \& {Bolton}(1986)}]{1986ApJS...61..419G}
{Gies}, D.~R., \& {Bolton}, C.~T. 1986, \apjs, 61, 419, \dodoi{10.1086/191118}

\bibitem[{{Gilliland} \& {Dupree}(1996)}]{1996ApJ...463L..29G}
{Gilliland}, R.~L., \& {Dupree}, A.~K. 1996, \apjl, 463, L29,
  \dodoi{10.1086/310043}

\bibitem[{{Grunhut} {et~al.}(2012){Grunhut}, {Wade}, {Sundqvist}, {ud-Doula},
  {Neiner}, {Ignace}, {Marcolino}, {Rivinius}, {Fullerton}, {Kaper},
  {Mauclaire}, {Buil}, {Garrel}, {Ribeiro}, \& {Ubaud}}]{2012MNRAS.426.2208G}
{Grunhut}, J.~H., {Wade}, G.~A., {Sundqvist}, J.~O., {et~al.} 2012, \mnras,
  426, 2208, \dodoi{10.1111/j.1365-2966.2012.21799.x}

\bibitem[{{Harper} {et~al.}(2008){Harper}, {Brown}, \&
  {Guinan}}]{2008AJ....135.1430H}
{Harper}, G.~M., {Brown}, A., \& {Guinan}, E.~F. 2008, \aj, 135, 1430,
  \dodoi{10.1088/0004-6256/135/4/1430}

\bibitem[{{Harper} {et~al.}(2017){Harper}, {Brown}, {Guinan}, {O'Gorman},
  {Richards}, {Kervella}, \& {Decin}}]{2017AJ....154...11H}
{Harper}, G.~M., {Brown}, A., {Guinan}, E.~F., {et~al.} 2017, \aj, 154, 11,
  \dodoi{10.3847/1538-3881/aa6ff9}

\bibitem[{{Heger} {et~al.}(2003){Heger}, {Fryer}, {Woosley}, {Langer}, \&
  {Hartmann}}]{2003ApJ...591..288H}
{Heger}, A., {Fryer}, C.~L., {Woosley}, S.~E., {Langer}, N., \& {Hartmann},
  D.~H. 2003, \apj, 591, 288, \dodoi{10.1086/375341}

\bibitem[{{Heger} {et~al.}(2000){Heger}, {Langer}, \&
  {Woosley}}]{2000ApJ...528..368H}
{Heger}, A., {Langer}, N., \& {Woosley}, S.~E. 2000, \apj, 528, 368,
  \dodoi{10.1086/308158}

\bibitem[{{Heger} {et~al.}(2005){Heger}, {Woosley}, \&
  {Spruit}}]{2005ApJ...626..350H}
{Heger}, A., {Woosley}, S.~E., \& {Spruit}, H.~C. 2005, \apj, 626, 350,
  \dodoi{10.1086/429868}

\bibitem[{{Higgins} \& {Vink}(2019)}]{2019A&A...622A..50H}
{Higgins}, E.~R., \& {Vink}, J.~S. 2019, \aap, 622, A50,
  \dodoi{10.1051/0004-6361/201834123}

\bibitem[{Hunter(2007)}]{Hunter2007}
Hunter, J.~D. 2007, Computing In Science \& Engineering, 9, 90,
  \dodoi{10.1109/MCSE.2007.55}

\bibitem[{{Ivanova}(2002)}]{2002PhDT........25I}
{Ivanova}, N. 2002, PhD thesis, University of Oxford

\bibitem[{{Ivanova} \& {Podsiadlowski}(2003)}]{2003fthp.conf...19I}
{Ivanova}, N., \& {Podsiadlowski}, P. 2003, in From Twilight to Highlight: The
  Physics of Supernovae, ed. W.~{Hillebrandt} \& B.~{Leibundgut}, 19

\bibitem[{{Ivanova} {et~al.}(2002){Ivanova}, {Podsiadlowski}, \&
  {Spruit}}]{2002MNRAS.334..819I}
{Ivanova}, N., {Podsiadlowski}, P., \& {Spruit}, H. 2002, \mnras, 334, 819,
  \dodoi{10.1046/j.1365-8711.2002.05543.x}

\bibitem[{{Kadam} {et~al.}(2016){Kadam}, {Motl}, {Frank}, {Clayton}, \&
  {Marcello}}]{2016MNRAS.462.2237K}
{Kadam}, K., {Motl}, P.~M., {Frank}, J., {Clayton}, G.~C., \& {Marcello}, D.~C.
  2016, \mnras, 462, 2237, \dodoi{10.1093/mnras/stw1814}

\bibitem[{{Kervella} {et~al.}(2018){Kervella}, {Decin}, {Richards}, {Harper},
  {McDonald}, {O'Gorman}, {Montarg{\`e}s}, {Homan}, \&
  {Ohnaka}}]{2018A&A...609A..67K}
{Kervella}, P., {Decin}, L., {Richards}, A. M.~S., {et~al.} 2018, \aap, 609,
  A67, \dodoi{10.1051/0004-6361/201731761}

\bibitem[{{Kippenhahn} {et~al.}(1980){Kippenhahn}, {Ruschenplatt}, \&
  {Thomas}}]{1980A&A....91..175K}
{Kippenhahn}, R., {Ruschenplatt}, G., \& {Thomas}, H.-C. 1980, \aap, 91, 175

\bibitem[{{Kippenhahn} \& {Weigert}(1967)}]{1967ZA.....65..251K}
{Kippenhahn}, R., \& {Weigert}, A. 1967, \zap, 65, 251

\bibitem[{{Kroupa} {et~al.}(2018){Kroupa}, {Je{\v{r}}{\'a}bkov{\'a}},
  {Dinnbier}, {Beccari}, \& {Yan}}]{2018A&A...612A..74K}
{Kroupa}, P., {Je{\v{r}}{\'a}bkov{\'a}}, T., {Dinnbier}, F., {Beccari}, G., \&
  {Yan}, Z. 2018, \aap, 612, A74, \dodoi{10.1051/0004-6361/201732151}

\bibitem[{{Lambert} {et~al.}(1984){Lambert}, {Brown}, {Hinkle}, \&
  {Johnson}}]{1984ApJ...284..223L}
{Lambert}, D.~L., {Brown}, J.~A., {Hinkle}, K.~H., \& {Johnson}, H.~R. 1984,
  \apj, 284, 223, \dodoi{10.1086/162401}

\bibitem[{{Landau} \& {Lifshitz}(1959)}]{1959flme.book.....L}
{Landau}, L.~D., \& {Lifshitz}, E.~M. 1959, {Fluid mechanics}

\bibitem[{{Langer}(2012)}]{2012ARA&A..50..107L}
{Langer}, N. 2012, \araa, 50, 107, \dodoi{10.1146/annurev-astro-081811-125534}

\bibitem[{{Mackey} {et~al.}(2014){Mackey}, {Langer}, {Meyer}, {Gvaramadze},
  {Mohamed}, {Neilson}, \& {Mignone}}]{2014arXiv1406.0878M}
{Mackey}, J., {Langer}, N., {Meyer}, D.~M.-A., {et~al.} 2014, ArXiv e-prints.
\newblock \doarXiv{1406.0878}

\bibitem[{Maeder \& Meynet(2000)}]{rotation}
Maeder, A., \& Meynet, G. 2000, Annual Review of Astronomy and Astrophysics,
  38, 143, \dodoi{10.1146/annurev.astro.38.1.143}

\bibitem[{{Mason} {et~al.}(2009){Mason}, {Hartkopf}, {Gies}, {Henry}, \&
  {Helsel}}]{2009AJ....137.3358M}
{Mason}, B.~D., {Hartkopf}, W.~I., {Gies}, D.~R., {Henry}, T.~J., \& {Helsel},
  J.~W. 2009, \aj, 137, 3358, \dodoi{10.1088/0004-6256/137/2/3358}

\bibitem[{{Menon} \& {Heger}(2017)}]{2017MNRAS.469.4649M}
{Menon}, A., \& {Heger}, A. 2017, \mnras, 469, 4649.
\newblock \doarXiv{1703.04918}

\bibitem[{{Menon} {et~al.}(2019){Menon}, {Utrobin}, \&
  {Heger}}]{2019MNRAS.482..438M}
{Menon}, A., {Utrobin}, V., \& {Heger}, A. 2019, \mnras, 482, 438,
  \dodoi{10.1093/mnras/sty2647}

\bibitem[{{Meynet} {et~al.}(2013){Meynet}, {Haemmerl{\'e}}, {Ekstr{\"o}m},
  {Georgy}, {Groh}, \& {Maeder}}]{2013EAS....60...17M}
{Meynet}, G., {Haemmerl{\'e}}, L., {Ekstr{\"o}m}, S., {et~al.} 2013, in EAS
  Publications Series, Vol.~60, EAS Publications Series, ed. P.~{Kervella},
  T.~{Le Bertre}, \& G.~{Perrin}, 17--28

\bibitem[{{Mikkola}(1983)}]{1983MNRAS.203.1107M}
{Mikkola}, S. 1983, \mnras, 203, 1107, \dodoi{10.1093/mnras/203.4.1107}

\bibitem[{{Mohamed} {et~al.}(2012){Mohamed}, {Mackey}, \&
  {Langer}}]{2012A&A...541A...1M}
{Mohamed}, S., {Mackey}, J., \& {Langer}, N. 2012, \aap, 541, A1,
  \dodoi{10.1051/0004-6361/201118002}

\bibitem[{{Motl} {et~al.}(2007){Motl}, {Frank}, {Tohline}, \&
  {D'Souza}}]{2007ApJ...670.1314M}
{Motl}, P.~M., {Frank}, J., {Tohline}, J.~E., \& {D'Souza}, M.~C.~R. 2007,
  \apj, 670, 1314, \dodoi{10.1086/522076}

\bibitem[{{Nance} {et~al.}(2018){Nance}, {Sullivan}, {Diaz}, \&
  {Wheeler}}]{2018MNRAS.479..251N}
{Nance}, S., {Sullivan}, J.~M., {Diaz}, M., \& {Wheeler}, J.~C. 2018, \mnras,
  479, 251, \dodoi{10.1093/mnras/sty1418}

\bibitem[{{Noriega-Crespo} {et~al.}(1997){Noriega-Crespo}, {van Buren}, {Cao},
  \& {Dgani}}]{1997AJ....114..837N}
{Noriega-Crespo}, A., {van Buren}, D., {Cao}, Y., \& {Dgani}, R. 1997, \aj,
  114, 837, \dodoi{10.1086/118517}

\bibitem[{{Oh} \& {Kroupa}(2016)}]{2016A&A...590A.107O}
{Oh}, S., \& {Kroupa}, P. 2016, \aap, 590, A107,
  \dodoi{10.1051/0004-6361/201628233}

\bibitem[{{Paxton} {et~al.}(2011){Paxton}, {Bildsten}, {Dotter}, {Herwig},
  {Lesaffre}, \& {Timmes}}]{2011ApJS..192....3P}
{Paxton}, B., {Bildsten}, L., {Dotter}, A., {et~al.} 2011, \apjs, 192, 3,
  \dodoi{10.1088/0067-0049/192/1/3}

\bibitem[{{Paxton} {et~al.}(2013){Paxton}, {Cantiello}, {Arras}, {Bildsten},
  {Brown}, {Dotter}, {Mankovich}, {Montgomery}, {Stello}, {Timmes}, \&
  {Townsend}}]{2013ApJS..208....4P}
{Paxton}, B., {Cantiello}, M., {Arras}, P., {et~al.} 2013, \apjs, 208, 4,
  \dodoi{10.1088/0067-0049/208/1/4}

\bibitem[{{Paxton} {et~al.}(2015){Paxton}, {Marchant}, {Schwab}, {Bauer},
  {Bildsten}, {Cantiello}, {Dessart}, {Farmer}, {Hu}, {Langer}, {Townsend},
  {Townsley}, \& {Timmes}}]{2015ApJS..220...15P}
{Paxton}, B., {Marchant}, P., {Schwab}, J., {et~al.} 2015, \apjs, 220, 15,
  \dodoi{10.1088/0067-0049/220/1/15}

\bibitem[{{Paxton} {et~al.}(2018){Paxton}, {Schwab}, {Bauer}, {Bildsten},
  {Blinnikov}, {Duffell}, {Farmer}, {Goldberg}, {Marchant}, {Sorokina},
  {Thoul}, {Townsend}, \& {Timmes}}]{2018ApJS..234...34P}
{Paxton}, B., {Schwab}, J., {Bauer}, E.~B., {et~al.} 2018, \apjs, 234, 34,
  \dodoi{10.3847/1538-4365/aaa5a8}

\bibitem[{{Paxton} {et~al.}(2019){Paxton}, {Smolec}, {Schwab}, {Gautschy},
  {Bildsten}, {Cantiello}, {Dotter}, {Farmer}, {Goldberg}, {Jermyn}, {Kanbur},
  {Marchant}, {Thoul}, {Townsend}, {Wolf}, {Zhang}, \&
  {Timmes}}]{2019ApJS..243...10P}
{Paxton}, B., {Smolec}, R., {Schwab}, J., {et~al.} 2019, \apjs, 243, 10,
  \dodoi{10.3847/1538-4365/ab2241}

\bibitem[{{Podsiadlowski}(2010)}]{2010NewAR..54...39P}
{Podsiadlowski}, P. 2010, \nar, 54, 39, \dodoi{10.1016/j.newar.2010.09.023}

\bibitem[{{Podsiadlowski} {et~al.}(1992){Podsiadlowski}, {Joss}, \&
  {Hsu}}]{1992ApJ...391..246P}
{Podsiadlowski}, P., {Joss}, P.~C., \& {Hsu}, J.~J.~L. 1992, \apj, 391, 246,
  \dodoi{10.1086/171341}

\bibitem[{{Podsiadlowski} {et~al.}(1990){Podsiadlowski}, {Joss}, \&
  {Rappaport}}]{1990A&A...227L...9P}
{Podsiadlowski}, P., {Joss}, P.~C., \& {Rappaport}, S. 1990, \aap, 227, L9

\bibitem[{{Pols}(1994)}]{1994A&A...290..119P}
{Pols}, O.~R. 1994, \aap, 290, 119

\bibitem[{{Poveda}(1964)}]{1964Natur.202.1319P}
{Poveda}, A. 1964, \nat, 202, 1319, \dodoi{10.1038/2021319b0}

\bibitem[{{Ram{\'\i}rez-Agudelo} {et~al.}(2013){Ram{\'\i}rez-Agudelo},
  {Sim{\'o}n-D{\'\i}az}, {Sana}, {de Koter}, {Sab{\'\i}n-Sanjul{\'\i}an}, {de
  Mink}, {Dufton}, {Gr{\"a}fener}, {Evans}, {Herrero}, {Langer}, {Lennon},
  {Ma{\'\i}z Apell{\'a}niz}, {Markova}, {Najarro}, {Puls}, {Taylor}, \&
  {Vink}}]{2013A&A...560A..29R}
{Ram{\'\i}rez-Agudelo}, O.~H., {Sim{\'o}n-D{\'\i}az}, S., {Sana}, H., {et~al.}
  2013, \aap, 560, A29, \dodoi{10.1051/0004-6361/201321986}

\bibitem[{{Ram{\'\i}rez-Agudelo} {et~al.}(2015){Ram{\'\i}rez-Agudelo}, {Sana},
  {de Mink}, {H{\'e}nault-Brunet}, {de Koter}, {Langer}, {Tramper},
  {Gr{\"a}fener}, {Evans}, \& {Vink}}]{2015A&A...580A..92R}
{Ram{\'\i}rez-Agudelo}, O.~H., {Sana}, H., {de Mink}, S.~E., {et~al.} 2015,
  \aap, 580, A92, \dodoi{10.1051/0004-6361/201425424}

\bibitem[{{Renzo} {et~al.}(2019){Renzo}, {Zapartas}, {de Mink}, {G{\"o}tberg},
  {Justham}, {Farmer}, {Izzard}, {Toonen}, \& {Sana}}]{2019A&A...624A..66R}
{Renzo}, M., {Zapartas}, E., {de Mink}, S.~E., {et~al.} 2019, \aap, 624, A66,
  \dodoi{10.1051/0004-6361/201833297}

\bibitem[{{Sana} {et~al.}(2012){Sana}, {de Mink}, {de Koter}, {Langer},
  {Evans}, {Gieles}, {Gosset}, {Izzard}, {Le Bouquin}, \&
  {Schneider}}]{2012Sci...337..444S}
{Sana}, H., {de Mink}, S.~E., {de Koter}, A., {et~al.} 2012, Science, 337, 444,
  \dodoi{10.1126/science.1223344}

\bibitem[{{Schneider} {et~al.}(2016){Schneider}, {Podsiadlowski}, {Langer},
  {Castro}, \& {Fossati}}]{2016MNRAS.457.2355S}
{Schneider}, F.~R.~N., {Podsiadlowski}, P., {Langer}, N., {Castro}, N., \&
  {Fossati}, L. 2016, \mnras, 457, 2355, \dodoi{10.1093/mnras/stw148}

\bibitem[{{Smith}(2014)}]{2014ARA&A..52..487S}
{Smith}, N. 2014, \araa, 52, 487, \dodoi{10.1146/annurev-astro-081913-040025}

\bibitem[{{Spruit}(2002)}]{2002A&A...381..923S}
{Spruit}, H.~C. 2002, \aap, 381, 923, \dodoi{10.1051/0004-6361:20011465}

\bibitem[{{Tayar} {et~al.}(2015){Tayar}, {Ceillier},
  {Garc{\'\i}a-Hern{\'a}ndez}, {Troup}, {Mathur}, {Garc{\'\i}a}, {Zamora},
  {Johnson}, {Pinsonneault}, \& {M{\'e}sz{\'a}ros}}]{2015ApJ...807...82T}
{Tayar}, J., {Ceillier}, T., {Garc{\'\i}a-Hern{\'a}ndez}, D.~A., {et~al.} 2015,
  \apj, 807, 82, \dodoi{10.1088/0004-637X/807/1/82}

\bibitem[{{Timmes} {et~al.}(2000){Timmes}, {Hoffman}, \&
  {Woosley}}]{2000ApJS..129..377T}
{Timmes}, F.~X., {Hoffman}, R.~D., \& {Woosley}, S.~E. 2000, \apjs, 129, 377,
  \dodoi{10.1086/313407}

\bibitem[{{Timmes} \& {Swesty}(2000)}]{2000ApJS..126..501T}
{Timmes}, F.~X., \& {Swesty}, F.~D. 2000, \apjs, 126, 501,
  \dodoi{10.1086/313304}

\bibitem[{{Uitenbroek} {et~al.}(1998){Uitenbroek}, {Dupree}, \&
  {Gilliland}}]{1998AJ....116.2501U}
{Uitenbroek}, H., {Dupree}, A.~K., \& {Gilliland}, R.~L. 1998, \aj, 116, 2501,
  \dodoi{10.1086/300596}

\bibitem[{{van Loon} {et~al.}(2005){van Loon}, {Cioni}, {Zijlstra}, \&
  {Loup}}]{2005A&A...438..273V}
{van Loon}, J.~T., {Cioni}, M. R.~L., {Zijlstra}, A.~A., \& {Loup}, C. 2005,
  \aap, 438, 273, \dodoi{10.1051/0004-6361:20042555}

\bibitem[{{Vink} {et~al.}(2001){Vink}, {de Koter}, \&
  {Lamers}}]{2001A&A...369..574V}
{Vink}, J.~S., {de Koter}, A., \& {Lamers}, H.~J.~G.~L.~M. 2001, \aap, 369,
  574, \dodoi{10.1051/0004-6361:20010127}

\bibitem[{{Wang} {et~al.}(2019){Wang}, {Kroupa}, \&
  {Jerabkova}}]{2019MNRAS.484.1843W}
{Wang}, L., {Kroupa}, P., \& {Jerabkova}, T. 2019, \mnras, 484, 1843,
  \dodoi{10.1093/mnras/sty2232}

\bibitem[{{Wellstein} {et~al.}(2001){Wellstein}, {Langer}, \&
  {Braun}}]{2001A&A...369..939W}
{Wellstein}, S., {Langer}, N., \& {Braun}, H. 2001, \aap, 369, 939,
  \dodoi{10.1051/0004-6361:20010151}

\bibitem[{{Wheeler} {et~al.}(2017){Wheeler}, {Nance}, {Diaz}, {Smith},
  {Hickey}, {Zhou}, {Koutoulaki}, {Sullivan}, \&
  {Fowler}}]{2017MNRAS.465.2654W}
{Wheeler}, J.~C., {Nance}, S., {Diaz}, M., {et~al.} 2017, \mnras, 465, 2654,
  \dodoi{10.1093/mnras/stw2893}

\bibitem[{{Zahn}(1977)}]{1977A&A....57..383Z}
{Zahn}, J.-P. 1977, \aap, 57, 383

\end{thebibliography}



\end{document}